\documentclass[10pt,english,english,nofootinbib,notitlepage,superscriptaddress,aps,prd]{revtex4-1}

\usepackage{enumerate}
\usepackage{amsmath}
\usepackage{amsfonts}
\usepackage{amssymb}
\usepackage[utf8]{inputenc}
\usepackage[T1]{fontenc}
\usepackage{mathtools}
\usepackage{wasysym}
\usepackage{accents}
\usepackage[svgnames]{xcolor}
\usepackage[export]{adjustbox}
\usepackage[colorlinks=true, pdfstartview=FitV, linkcolor=blue, citecolor=red, urlcolor=magenta]{hyperref}
\usepackage{url}
\usepackage{bm}
\usepackage{tikz}
\usetikzlibrary{arrows,decorations.markings}
\usepackage{csquotes}
\usepackage{appendix}
\usepackage{caption}
\usepackage{graphicx}
\usepackage{float}
\usepackage{subcaption}
\newcommand{\sech}{\mathrm{sech} \,} 

\begin{document}
\newcommand{\Areia}{
\affiliation{Department of Chemistry and Physics, Federal University of Para\'iba, Rodovia BR 079 - Km 12, 58397-000 Areia-PB,  Brazil.}
}
\newcommand{\Lavras}{
\affiliation{Physics Department, Federal University of Lavras, Caixa Postal 3037, 37200-000 Lavras-MG, Brazil.}
}

\newcommand{\JP}{
\affiliation{Physics Department, Federal University of Para\'iba, Caixa Postal 5008, 58059-900, Jo\~ao Pessoa, PB, Brazil.}
}

\title{Massless Dirac Perturbations in a Consistent Model of Loop Quantum Gravity Black Hole: Quasinormal Modes and Particle Emission Rates}

\author{Saulo Albuquerque}\email{saulosoaresfisica@gmail.com}
\JP \affiliation{Theoretical Astrophysics, Institute for Astronomy and Astrophysics, University of T\"{u}bingen, 72076 T\"{u}bingen, Germany.}

\author{Iarley P. Lobo}\email{lobofisica@gmail.com}
\Areia
\Lavras

\author{Valdir B. Bezerra}\email{valdir@fisica.ufpb.br}
\JP

\date{\today}

\begin{abstract}
We consider perturbations of the massless Dirac field in the background of a black hole solution found by Bodendorfer, Mele, and M\"{u}nch (BMM), using a polymerization technique that furnishes contributions inspired by Loop Quantum Gravity (LQG) Theory. Using the sixth order WKB method, we analyzed its quasinormal modes for several modes, multipole numbers and the two classes of BMM black holes. We also considered the potential that governs these perturbations to analyze the bound on the Greybody Factor (GF) due the emission rates of particles. As results, we found that the Loop Quantum Gravity parameters are responsible for raising the potential and the real and imaginary parts of the quasinormal frequencies and decrease the bound on the Greybody Factor for the two classes of black holes (with more prominent effects for the de-amplification case, which is compatible with previous analyses done for other fields).
\end{abstract}

\maketitle

%%%%%%%%%%%%%%%%%%%%%%%%%%%%%%%%%%%%%%%%%%%%%%%%%%%%%%%%%%%%%%%%%%%%%%%%%%%%%%%%%%%%%%%%%%%%%%%%%%%%%%

\section{Introduction}

One of the most important and widely studied approaches that aim to describe a quantized spacetime is the Loop Quantum Gravity (LQG) Theory \cite{Rova,Ashtekar_2004,https://doi.org/10.48550/arxiv.hep-th/0303185,Thiemann_2003,Ashtekar_2013}. % Loop quantum gravity (LQG) is an attempt to quantize gravity by itself, without the need to unify it with other fundamental interactions \cite{Gambini:2011zz}. This theory directly quantizes the classical gravitational theory (General Relativity Theory). 
It introduces the concept of quantum geometry in the sense that the spacetime itself is made of some fundamental building blocks \cite{Gambini:2011zz}, called spin-networks. The geometrical operators in the theory, such as volume and area, are now quantum operators \cite{Rovelli_1995} which have a discrete spectrum at the Planck scale. %The area operator, for instance, has the so-called minimum area-gap \cite{Rovelli_1995,Ashtekar_1997}.
%Those spin-networks have a scale above the order of the Planck length.

Just like other theoretical proposals to solve the quantum gravity/quantum spacetime problem, the LQG Theory faces a common problem that seems to affect all other quantum gravity theories. And this problem is the difficulty to test their predictions. In other words, the major issue faced by these approaches is the absence of experiments with the proper sensitiveness for the scale of phenomena that they predict. 
%So if we do not still possess the capacity to produce properly controlled and repeatable experiments to address the problem of testing quantum spacetime or quantum gravity theories,
%what is left for us to do regarding the quantum-gravity problem in a scientific way? What else can we do to reach the quantum spacetime realm? If we want to investigate this problem in a scientific way, this problem must be treated like any other problem in science: trying to find the guidance of experimental facts \cite{camelia}. 
Devoted to this problem, a large number of researchers in the last few years have gathered their efforts in attempts to propose
phenomenological tests for the quantum gravity/quantum spacetime problem \cite{camelia,Addazi:2021xuf}.

However, when it comes to the LQG, the difficulty to manage its complex formalism has shown to be a huge problem, mainly in what concerns finding physical predictions. Among many difficulties, the most notable one is the ``problem of the classical limit'' \cite{camelia}. This problem makes very challenging to find a way for LQG predictions for the quasi-Minkowski regime, which is usually where most opportunities for phenomenology come from. To address this problem, LQG phenomenologists have been focused on trying to infer, from the general structure of the theory, some candidates for effective models of the theory. 

%ultima sentença mais importante

%Resumir parte anterior a isso

The difficulty and complexity of the LQG formalism is also present when we consider the description of black holes, in this context. This difficulty obstructs the way towards obtaining physical predictions from this theory in the investigation of those astrophysical compact objects. This fact has motivated LQG phenomenologists to develop and use effective models in the context of black holes as well \cite{Modesto_2004,Modesto_2010,modesto2008black,bodendorfer2019effective}.

The general idea behind constructing those Loop Quantum Gravity black hole effective models (LQGBHs) is that the non-perturbative quantum geometry corrections, introduced by hand, modify Einstein equations. In practice, for effective versions of LQG, it is introduced holonomy modifications. Those transformations are quantum corrections resulting from the area-gap structure of the Hilbert space in LQG \cite{ashtekar1995quantization,ashtekar2003mathematical,ashtekar1997quantum}.

Therefore, the main focus of this framework here will be the application of some of those properties of spacetime quantization motivated by LQG Theory to investigate its signature on phenomenological features of black holes. The effects of spacetime quantization on the behavior of those astrophysical objects shall be investigated by means of an effective black hole model inspired by the LQG Theory \cite{bodendorfer2019effective}. The black holes' phenomenological features we shall investigate in this framework here are the ``ringdown frequencies'' of the massless Dirac perturbations on those black hole spacetimes, and the emission rate and ``greybody factor'' for the Dirac radiation produced by those compact objects. 

The previously mentioned ``ringdown frequencies'' of a field perturbation on a black hole spacetime are the characteristic frequencies of damped oscillations of this field in response to a linear perturbation. They dominate the signal of the gravitational waves at the ringdown stage \cite{cardosothesis,Vishveshwara:1970zz,cardoso2,cardoso3,cardoso4} and represent an intrinsic characteristic of the black holes. Those frequencies are often called 'quasinormal modes frequencies', and they form a discrete spectrum of complex frequencies. 

Mathematically, the quasinormal modes frequencies depend solely on the basic three parameters of the black holes (mass, charge, angular momentum). However, if there are any additional parameters that describe the black hole spacetime, such as quantum correction parameters (like the cases we study in this framework), for example, those parameters will also leave their signatures in the QNMs spectrum. This signature is exactly what we are willing to find. If we can detach this signature, we will be able to study the effects the quantum corrections proposed by the Loop Quantum Gravity effective model \cite{bodendorfer2019effective} provide on the behavior of the black holes. 

Even though the evolution of other test fields (such as the Dirac field) at a black hole spacetime background seems less related to the gravitational wave signals, it still might provide us with interesting insights about some properties of those quantum corrected black holes, such as their stability. This is our motivation to investigate the Dirac quasinormal modes of the black hole solution found by \cite{bodendorfer2019effective} which we call here as BMM black hole (which was also systematically studied in \cite{Gan:2020dkb}). A previous work in literature has considered the calculation of these quasinormal modes for the scalar, electromagnetic and gravitational field of this black hole model \cite{Bouhmadi-Lopez:2020oia}. So it was natural for us to consider the generalization of those results to the case of massless spin $1/2$ perturbing fields as well.

A few works have been devoted to finding those QNMs for the quantum corrected BMM black hole \cite{Bouhmadi-Lopez:2020oia}, and also for Modesto's model  \cite{Chen_2011,Santos,Cruz:2020emz}. The shadow and QNMs of rotating self-dual black holes \cite{modesto2008black} have also been studied \cite{qnmsrotatinglqg}. As we have previously mentioned, the only case that had not yet been considered was the Dirac perturbation. So we decided to investigate that specific case in this framework. As usual, for every new result we obtained, we compared it with its Schwarzschild analog. For the goals of this part of the paper to be accomplished, we used the 6th order WKB method \cite{Konoplya_2003}.

Another goal of this framework is to find the Dirac Particles' Emission Rate and the Greybody Factor of the Dirac radiation from this LQG black hole \cite{bodendorfer2019effective} and to study the signature of the LQG corrections in their behavior. The Greybody Factor (GF) is the quantity we define to measure the difference between the spectrum of radiation observed by an asymptotic observer and the black body's ideal radiation spectrum.

%Hawking showed in 1975 \cite{hawking1975particle} that black holes actually emit thermal radiation. According to Hawking, the thermal spectrum of radiation from a black hole matches the thermal spectrum from a common black body with a temperature that equals exactly the black hole's surface gravity. The only difference between the thermal spectrum of radiation from a black hole and a common black body is explained by the fact that, for black holes, before reaching an observer at spatial infinity and being measured, this thermal radiation must travel through a curved spacetime geometry. As a result, the surrounding spacetime plays the role of a potential barrier, causing a variation from the ideal black body radiation spectrum to the radiation spectrum actually observed by the asymptotic observer \cite{SAKALLI_2022}. The quantity we define to measure the difference between the spectrum of radiation observed by the asymptotic observer and the black body's radiation spectrum is the Grey-Body Factor (GF).

The Hawking radiation spectrum and its GF are very sensitive to modifications of general relativity. So they provide, at least in principle, an important source of physical consequences for the modifications that might be considered in the formulation of the black hole spacetime. If a quantum correction is considered when one formulates the effective metric of a black hole spacetime, this correction will leave a signature on the Hawking radiation spectrum and on its Greybody Gactor.

%Therefore, it is another goal of this framework to find the Greybody Factor of the Dirac radiation from this LQG black hole and to study the signature of the Loop Quantum Gravity corrections in its behavior.

Studying the physical consequences of LQG corrections on the GF of the Dirac field of this polymerised black hole \cite{bodendorfer2019effective} required us to derive and solve the equation of motion of this perturbing field on the background of that LQG black hole spacetime. In other words, we had to derive and solve the Dirac equation. This was a similar problem to the one that we faced when dealing with the quasinormal modes. However, for this specific part we used a particular technique, namely the 'Bounding Bogoliubov Coefficients' method \cite{Visser_1999, Boonserm_2008,Shankaranarayanan_2003,Boonserm_2010}.

This work is divided in the following way: in section \ref{sec:bmm} we give a review of the LQG effective model we will use as the theoretical background for our investigation \cite{bodendorfer2019effective}; in section \ref{sec:modes}, we formulate the Dirac equation at the neighbourhood of this LQG black hole, and we use the 6th order WKB method to find the quasinormal frequencies from this Dirac equation; and finally, in section \ref{sec:grey}, we recall the Dirac equation we formulated to evaluate the GF of the Dirac radiation emerging from this quantum corrected black hole. Our main goal for the final two sections will be to detach the signature of the Loop Quantum Gravity corrections, carried out by the effective model used here \cite{Bodendorfer_2019}, in the quasinormal modes spectra of the Dirac massless field and in the Greybody Gactor of the Dirac radiation.  Then, we conclude in section \ref{sec:final}.

\section{A consistent model of non-singular Schwarzschild black hole in loop quantum gravity - The BMM Black Hole}\label{sec:bmm}

An example of an effective model that gave us some hints about how we could study black hole
physics in the context of Loop Quantum Gravity came from cosmology, more specifically, from
Loop Quantum Cosmology  (LQC) \cite{Ashtekar_2011,oriti2017bouncing,ashtekar2017loop,ashtekar2006quantum,ashtekar2015loop}. The simplest case is provided by the LQC description of Friedmann-Lema\^itre-Robertson-Walker (FLRW)
cosmological spacetimes. In these LQC cosmology models, quantum geometry effects lead to an
effective spacetime where the `Big Bang' singularity was replaced by a `Big Bounce', i.e. a
quantum regime which interpolates between a contracting and an expanding branch. The root of
the effective quantum theory model was a phase space regularisation, usually called
polymerisation in the LQG literature.

In the historical endeavor towards the study of black holes in a Loop Quantum Gravity
formulation, the first attempt was made by Modesto \cite{modesto2008black,Modesto_2004}. He employed semiclassical techniques in the minisuperspace quantization
scheme \cite{constraintsselfdualblackhole}. One of the main successes accomplished by these first models
was the possibility to address the problem of the black hole singularity \cite{Modesto_2004}.

That technique was generalized to the approach of polymerisation for black hole spacetime
solutions \cite{Modesto_2010}. In a subsequent work by Modesto\cite{Modesto_2010}, it was tried to improve
the semiclassical analysis by introducing a simple modification to the holonomic version of the
Hamiltonian constraint. The main result is that the minimum area \cite{rovelli1990loop} of full LQG is the fundamental ingredient to solving the black hole space-time singularity problem in $r=0$. The ${\mathbb S}^2$ sphere
bounces on the minimum area of LQG and the singularity disappears. Those new defined black
hole models for a quantum corrected Schwarzschild spacetime were characterized by a
polymeric parameter which measures the quantum correction to the classical spacetime.
Therefore, large enough black holes would then be described by a classical black hole metric
encoding the quantum corrections in a suitable way. The procedure was later generalized for
spinning black holes \cite{Caravelli_2010}.

%The general idea behind constructing these effective models in LQG is that the non-perturbative quantum geometry corrections, introduced by hand, modify Einstein equations. In practice, for effective versions of LQG, those holonomy modifications that we mentioned before are quantum corrections resulting from the area-gap structure of the Hilbert Space in LQG \cite{ashtekar1995quantization,ashtekar2003mathematical,ashtekar1997quantum}.

That method of introducing those holonomy modifications in effective models, called
polymerisation technique, is achieved in the following way: starting from the canonically
conjugate phase space variables $(q,p)$, describing the geometry of the minisuperspace model
under consideration, we replace the conjugate momenta $p$ in the phase space with their
polymeriezed version $\sin{(\lambda p)}/\lambda$, where $\lambda$ is the quantum parameter related to the area-gap, called ``polymerisation scale''. This parameter controls the onset of quantum effects. That
trigonometric function is not arbitrary, rather it is a result of having matrix elements of $SU(2)$
holonomies evaluated along a loop. The curvature of spacetime, calculated in terms of these
holonomies, is naturally regularized as a result of using those bounded functions. As a
consequence of this method, a transition surface inside the black hole
replaces the classical singularity\cite{modesto2008black,Modesto_2010, bodendorfer2019effective}.

For the effective black hole model in LQG proposed by Modesto \cite{modesto2008black}, some criticisms arose regarding fundamental inconsistencies that the model suffered from \cite{Brahma_2018, Bodendorfer_2019,bojowaldrage,towconsbtwhbfmc,Achour_2020,Ben_Achour_2020}.
Aware of those inconsistencies, Bodendorfer, Mele and Munch (BMM) proposed a new effective model \cite{bodendorfer2019effective} of a quantum corrected black hole by using the same method of polymerisation of canonical phase space variables by holonomy transformations. This time, however, instead of
the $SU(2)$ connections and the conjugate momenta, the BMM model is based on the
polymerisation of a new set of canonical phase space variables. Three important features
characterized the resulting effective model:

\begin{itemize}
    \item The spacetime singularity of general relativity is replaced with a spacelike transition
surface, separating infinite pairs of trapped and anti-trapped regions.
    \item Quantum effects become relevant at a unique mass-independent curvature scale, while
they become negligible in the low curvature region near the horizon.
    \item The solution recovers the Schwarzschild black hole solution near the event horizon, and
it is also asymptotically flat.
\end{itemize}

In this framework, we will be focused on that proposal, called the BMM model\cite{bodendorfer2019effective} of a
quantum corrected black hole in Loop Quantum Gravity. We shall compare its predictions with the classical Schwarzschild predictions in order to investigate the signature of the LQG corrections in the behavior of the quasinormal modes of the Dirac field and in the behavior of the Hawking emission rate and the Greybody Factor of the Dirac radiation.

The spacetime metric for the effective model of the quantum corrected BMM black hole is given by the line element below (followins the notation of \cite{Bouhmadi-Lopez:2020oia}):

\begin{equation}\label{metric}
    ds^2= \frac{-4a(b)A^2B^{2/3}}{\lambda_2^2}d\tau^2+\frac{\lambda_2^2}{4a(b)}\left(1+\frac{1}{X(b)^2}\right)^2\left(\frac{dX}{db}\right)^2db^2+b^2d\Omega_2^2\, ,
\end{equation}
where $b$ is the radial coordinate, $d\Omega_2$ is the line element of the two-dimensional unit sphere ${\mathbb S}^2$, and $\lambda_2$ is a quantum gravity parameter of the polymerization, whose product has cubic dimensions of mass, and are such that when they go to zero, the solution reduces to the Schwarzschild spacetime. The functions of the metric are
\begin{equation}
    a(b)\doteq a(X(b))=\lambda_2^2\left(\frac{X^2+1}{2X}\right)^2\left(1-\frac{3CD}{2\lambda_2}\frac{2X}{X^2+1}\right)\frac{1}{b^2}\, ,
\end{equation}
where
\begin{equation}\label{x1}
    X(b)^3=\frac{b^3}{2A^3B}\pm \frac{1}{2}\sqrt{\frac{b^6}{(A^3B)^2}-\frac{4}{B}}\, ,
\end{equation}
for the following integration constants
\begin{align}
    &A\, B^{1/3}=\left[\frac{\lambda_1\lambda_2M_{BH}}{2}\left(\frac{M_{BH}}{M_{WH}}\right)^{3/2}\right]^{1/4}\, ,\\
    &B=\left(\frac{M_{BH}}{M_{WH}}\right)^3\, ,\\
    &C\, D=\frac{2}{\lambda_1}\left[\frac{2}{3}\left(\frac{\lambda_1\lambda_2}{3}\right)^3M_{BH}^3\left(\frac{M_{WH}}{M_{BH}}\right)^{3/2}\right]^{1/4}.
\end{align}

Notice from the above equations that this solution is actually described by two quantum gravity parameters, $\lambda_1$ and $\lambda_2$, such that when they are zero, we recover results of General Relativity. This solution also presents two Dirac observables, i.e., that are constant on-shell quantities, which are identified as $M_{BH}$ and $M_{WH}$. To interpret this solution, we rely on the properties of the function \eqref{x1}. Notice that it has two roots that behave very differently at an asymptotic region $b\rightarrow \infty$. When $b\rightarrow \infty$, we have $X\rightarrow \infty$ for the ``$+$'' root, which actually leads to the asymptotic line element
\begin{equation}
    ds_+^2\approx -\left(1-\frac{2M_{BH}}{b}\right)d\tau^2+\frac{db^2}{1-\frac{2M_{BH}}{b}}+b^2 d\Omega_2^2\, ,
\end{equation}
meaning that the spacetime approaches the one of a Schwarzschild black hole with mass $M_{BH}$. On the other hand, the negative root presents a different asymptotic behavior, now as $X\rightarrow 0_+$ when $b\rightarrow \infty$, which furnishes the following line element
\begin{equation}
    ds^2_{-}\approx -B^{2/3}\left(1-\frac{2M_{WH}}{b}\right)d\tau^2+\frac{db^2}{1-\frac{2M_{WH}}{b}}+b^2 d\Omega_2^2\, ,
\end{equation}
which means that after a time rescaling, we get, approximately, a Schwarzschild black hole with mass $M_{WH}$. These two braches are connected when the square root of \eqref{x1} vanishes and the radius attains the minimum value
\begin{equation}
    b_m=\left[2(\lambda_1\lambda_2)^3M_{BH}^3\left(\frac{M_{WH}}{M_{BH}}\right)^{3/2}\right]^{1/12}\, .
\end{equation}

This minimum means that the derivative $dX/db|_{b_m}\rightarrow \infty$, leading the divergence of the spatial metric function $g_{bb}=\lambda_2^2\left(1+X(b)^{-2}\right)^2\left(dX/db\right)^2/(4a)$ while the time function of the metric $g_{\tau\tau}=-4aA^2B^{2/3}/\lambda_2^2$ vanishes. This means that the singularity of this spacetime is removed due to quantum gravitational effects and we have, in fact, a connection of a black hole region with mass $M_{BH}$ to a white hole region with mass $M_{WH}$. This can be seen from the Penrose diagrams of this spacetime calculated in \cite{bodendorfer2019effective}. For this reason, this solution actually presents four free parameters: $\lambda_{1}$, $\lambda_2$, $M_{BH}$ and $M_{WH}$.
\par
For applications in phenomenology, it is necessary that one identifies a curvature scale from which quantum gravity effects become relevant. In order to have a unique such a scale, it is necessary to relate the mass scales $M_{BH}$ and $M_{WH}$, for instance, by a relation of the kind
\begin{equation}
    M_{WH}=M_{BH}\left(\frac{M_{BH}}{m}\right)^{\beta-1}\, ,
\end{equation}
where $m$ is a fiducial mass scale, which is assumed to be smaller than the masses of the black and white holes. The uniqueness of the curvature scale for quantum gravity leads to the following possibilities for the parameter $\beta$:
\begin{equation}
    \beta=5/3 \qquad \text{(Mass amplification)}
\end{equation}
leads to a phenomenon called mass amplification, due to the fact that when an observer travels from the black hole region to the white hole one, the noticed white hole presents a mass larger than the connected black hole, i.e., $M_{WH}>M_{BH}$. On the other hand, if
\begin{equation}
    \beta=3/5 \qquad \text{(Mass de-amplification)}
\end{equation}
an observer experiences a de-amplification in this transition, since $M_{WH}<M_{BH}$. For this reason, once fixing the fiducial mass scale $m$, we can substitute one of the masses (for instance $M_{WH}$) by the parameter $\beta$ and describe our quantities is terms of $\lambda_1$, $\lambda_2$, $M_{BH}$ and $\beta$.

\section{Quasinormal modes for massless spin $1/2$ particles}\label{sec:modes}

Typically, the quasinormal modes of a black hole are deduced by solving a family of master
wave equations, which are derived either by perturbing the spacetime metric directly, or by
considering the field equations of some test fields around the black hole spacetime (in other
words, considering the field equations in the curved spacetime of the proposed black hole). The
first case corresponds to the gravitational perturbations of the black hole, which are categorized
into axial and polar perturbations, and which generate the gravitational waves at the ringdown
stage. The second one, however, describes the evolution of the test fields in the background of this
black hole spacetime. Examples of test fields that may be considered in the background of a
black hole spacetime are the Dirac field, the scalar field, the electromagnetic field, and the
Proca field. To formulate those master equations, we write their respective field equations at
the black hole spacetime background. In this specific case considered in the present paper, we shall be focused on formulating the Dirac equation on the spacetime surrounding the BMM black hole \cite{Bodendorfer_2019}. In order to do that, we need to find, firstly, how the Dirac equation at the neighbourhood of this black hole is written. 

\subsection{The field equations for massless spin 1/2 fields at the vicinity of the BMM black hole}

In a generally curved spacetime, the Dirac equation is written as:

\begin{equation}
	(i\gamma^{\mu}(x)\nabla_{\mu}-\mu_{*})\Psi(x)=0\label{diracequation},
\end{equation} 

\noindent where the covariant derivative reads

\begin{equation}
	\nabla_{\mu}=\partial_{\mu}-\Gamma_{\mu}\label{nabla}.
\end{equation}

The $\Gamma_{\mu}$ are called spinorial connection symbols, and are given by

\begin{equation}
     \Gamma_{\mu}=\frac{1}{8}\gamma_{(b)(a)(c)}e_{\mu}^{(c)}[\gamma^{a},\gamma^{b}],
\end{equation}

\noindent where $\gamma_{(b)(a)(c)}$ are the Ricci coefficients and the $\gamma^{\mu}$ matrices are defined in a general curved spacetime as

\begin{equation}
	\gamma^{\mu}=e^{\mu}_{(a)}(x)\gamma^{(a)}\label{curvedgamma},
\end{equation}

\noindent with $\gamma^{(a)}$ being the constant Dirac matrices and $e^{\mu}_{(a)}(x)$ are the spacetime tetrads.

In order for the Dirac equation above remain consistent with the 2-spinor form of the Dirac equation as considered by Chandrasekhar \cite{chandrasekhar1976solution,page1976dirac}, we shall write the four component Dirac spinor as \cite{Teukolsky}:

\begin{equation}\label{2spinors}
	\Psi(x)=\begin{bmatrix}
	\chi^{A}(x) \\
	\eta_{B}(x)
	\end{bmatrix},
\end{equation}
\noindent where $\chi^{A}$ and $\eta_{B}$ are two-dimensional spinors.

In accordance with  Chandrasekhar \cite{chandrasekhar1976solution}, the Dirac equation is separable in the Kerr black hole space time, and thus, the following 2-spinor form of the Dirac equation can be obtained:

\begin{align}
	\nabla_{AB'}\chi^{A}+i\mu_{*}\bar{\eta}_{B'}=0,\\
    \nabla_{AB'}\eta^{A}+i\mu_{*}\bar{\chi}_{B'}=0,
\end{align}

\noindent where $\nabla_{AB'}=\sigma^{\mu}_{AB'}\nabla_{\mu}$ and $\mu_{*}$ is the particle mass. The symbols $\sigma^{\mu}_{AB'}$ are specified by the specific choice of representation for the $\gamma^{\mu}$ matrices. 

For a massless Dirac field, we are left with:

\begin{align}
	\nabla_{AB'}\chi^{A}=0, \label{20}\\
    \nabla_{AB'}\eta^{A}=0.
\end{align}

In the Newman-Penrose formalism \cite{NP}, the equations \eqref{20} become \cite{Teukolsky}:

\begin{align}
    &(D+\epsilon-\rho)\chi_{1}-(\bar{\delta}+\pi-\alpha)\chi_{0}=0, \label{conditions1} \\
    &(\Delta + \mu -\gamma)\chi_{0}-(\delta + \beta -\tau)\chi_{1}=0, \label{conditions2}
\end{align}

\noindent where

\begin{equation}
    D \equiv l^{a}\nabla_{a}, \qquad \Delta \equiv n^{a}\nabla_{a}, \qquad \delta \equiv m^{a}\nabla_{a}, \qquad \bar{\delta} \equiv \bar{m}^{a}\nabla_{a} \label{DDeltadeltadelta},
\end{equation}

\noindent and

\begin{align}
    \epsilon = (\gamma_{211} + \gamma_{341})/2 ,\qquad \rho = \gamma_{314},\qquad \pi = \gamma_{241},\qquad \alpha = (\gamma_{214} + \gamma_{344})/2, \\ 
    \mu=\gamma_{243},\qquad \gamma = (\gamma_{212} + \gamma_{342})/2, \qquad
    \tau = \gamma_{312}.
\end{align}

For general spherically symmetric static metrics which constitute a subset of Petrov type D metrics \cite{Arbey:2021jif} with the following general form:

\begin{equation}\label{metricFG}
    ds^2=-G(b)d\tau^2+\frac{1}{F(b)}db^2+b^2 d\Omega_{2}^2,
\end{equation}

\noindent we have that an appropriate choice for the null tetrad is \cite{Arbey:2021jif}:

\begin{align}
    l^{\mu}=\left(\frac{1}{G},\sqrt{\frac{F}{G}},0,0 \right) ,\qquad n^{\mu}=\left(\frac{1}{2},-\frac{\sqrt{FG}}{2},0,0\right), \label{ln}\\
    m^{\mu}=\left(0,0,\frac{1}{b\sqrt{2}},\frac{i}{b \sin{\theta}\sqrt{2}} \right), \qquad
    \bar{m}^{\mu}=\left(0,0,\frac{1}{b\sqrt{2}},-\frac{i}{b \sin{\theta}\sqrt{2}} \right) \label{mm}.
\end{align}

With those definitions, we have that the only nonvanishing components for the Ricci coefficients are real and given by:

\begin{align}
    \rho =-\frac{1}{b}\sqrt{\frac{F}{G}},\qquad
    \beta=-\alpha = \frac{\cot{\theta}}{2\sqrt{2}b}, \qquad
    \mu=-\frac{1}{2b}\sqrt{FG},\qquad
    \gamma =\frac{G^{'}}{4}\sqrt{\frac{F}{G}}  \label{rhoalfabetamugamma}.
\end{align}

For a general spherically symmetric static metrics such as the one in Eq. \eqref{metricFG}, we can decouple the set of differential equations \eqref{conditions1}, \eqref{conditions2} to produce a single differential equation for $\chi_{0}$ only (which is actually the equation of motion for a massless Dirac field):

\begin{equation}
    [(D-2\rho)(\Delta-\gamma+\mu)-(\delta-\alpha)(\bar{\delta}-\alpha)]\chi_{0}=0 \label{equationxo}.
\end{equation}

Now, using the definitions above for the null tetrad, given by Eqs. \eqref{ln},\eqref{mm}, the results found in Eq. \eqref{rhoalfabetamugamma}, and applying it all to Eq. \eqref{equationxo}, we finally obtain for the massless spin $1/2$ field the following explicit equation: 

\begin{align}
    &-\partial_{t}^2\chi_{0}+\frac{1}{2}\sqrt{\frac{F}{G}}\left(G'-\frac{2G}{b} \right)\partial_{t}\chi_{0}+\left(\frac{1}{\sin{\theta}}\partial_{\theta}(\sin{\theta}\partial_{\theta})+(\csc{\theta})^2\partial_{\phi}^2+\frac{i\cot{\theta}}{\sin{\theta}}\partial_{\phi}+\frac{1}{2}-\frac{1}{4}(\cot{\theta})^2\right)\chi_{0} \nonumber\\
    &+\left(\frac{FG''}{2}+\frac{FG}{b^2}-\frac{FG'^2}{4G}+\frac{F'G'}{4}+\frac{F'G}{2b}+\frac{3FG'}{2b}\right)\chi_{0}+\frac{1}{b^2}\sqrt{\frac{F}{b^2}}\left(\sqrt{b^2F}Gb^2 \chi_{0}'\right)'=0.
\end{align}

Now, using the ansatz:

\begin{equation}
    \chi_{0}=\Phi(b) S_{l,m}(\theta,\phi) e^{-i \omega t},
\end{equation}

\noindent the radial and angular parts decouple, and the angular part yields:

\begin{equation}
   \left(\frac{1}{\sin{\theta}}  \partial_{\theta}(\sin{\theta}\partial_{\theta})+(\csc{\theta})^2\partial_{\phi}^2 +\frac{i\cot{\theta}}{\sin{\theta}}\partial_{\phi} +\frac{1}{2}-\frac{1}{4}(\cot{\theta})^2 +\lambda_{l}\right)S_{l,m}(\theta,\phi)=0,
\end{equation}

\noindent where $\lambda_{l}$ is the decoupling constant.

Meanwhile, the radial part of the massless spin $1/2$ field equation becomes:

\begin{align}
    &\left(\omega^2 +i\omega \frac{1}{2}\sqrt{\frac{F}{G}}\left(\frac{2G}{b}-G'\right) + \frac{FG''}{2}+\frac{FG}{b^2}-\frac{FG'^2}{4G}+\frac{F'G'}{4}+\frac{F'G}{2b}+\frac{3FG'}{2b}-\frac{G(\lambda_{l}^{1/2}+1)}{b^2}\right)\Phi \nonumber\\
    &+\frac{1}{b^2}\sqrt{\frac{F}{b^2}}(\sqrt{b^2F}Gb^2 \Phi')'=0 \label{radialpart}.
\end{align}

This is the radial Teukolsky equation for a massless spin $1/2$ field. It is now convenient to define a generalized tortoise coordinate $b^{*}$ as:

\begin{equation}
    \frac{db^{*}}{db}=\frac{1}{\sqrt{FG}}.
\end{equation}

Having that definition for the tortoise coordinate and after some algebraic steps, the equation \eqref{radialpart} above can be transformed into the following Schr\"odinger-like wave equations for some certain wave functions $\psi(b(b^{*}))$:

\begin{equation}
    \partial_{*}^2 \psi +[\omega^2-V_{\pm}(b(b^{*}))]\psi=0 ,\label{masterequation}
\end{equation}

\noindent where $\partial_{*}$ denotes the derivative with respect to the tortoise coordinate $b^{*}$, and the short-ranged potentials $V_{\pm}(b(b^{*}))$ for the massless spin $1/2$ field are given by the following formulas for a general spherically symmetric static spacetime \cite{Arbey:2021jif}:

\begin{align}
    V_{\pm}(b(b^{*}))&=(l(l+1)+1/4)\frac{G}{b^2}\pm \sqrt{l(l+1)+1/4}\partial_{*}\left(\sqrt{\frac{G}{b^2}} \right)  \nonumber\\
    &=(l(l+1)+1/4)\frac{G}{b^2}\pm \sqrt{l(l+1)+1/4}\sqrt{FG}\left(\sqrt{\frac{G}{b^2}} \right)'.\label{potential}
\end{align}

By imposing that the spacetime metric is the one provided by Eq.(\ref{metric}), we find that the short-ranged potentials $V_{\pm}(b(b^{*}))$ in Eq. \eqref{potential} for a massless Dirac field at the vicinity of a polymer Schwarzschild black hole inspired by LQG \cite{Bodendorfer_2019} are given by:

\begin{equation}
    V_{\pm}(b)=\frac{4}{\lambda_{2}^2}\left[\frac{\lambda_{1}\lambda_{2}M_{BH}}{2} \left(\frac{M_{BH}}{M_{WH}} \right)^{\frac{3}{2}}\right]^{\frac{1}{2}}\left[(l(l+1)+1/4)\frac{a(b)}{b^2}\pm \sqrt{l(l+1)+1/4}\left(\frac{2}{\lambda_{2}}\right)\frac{a(b)}{\left(1+\frac{1}{X^2(b)}\right)\frac{dX}{db}}\left(\frac{\sqrt{a(b)}}{b} \right)'\right]. \label{potentialbmm}
\end{equation}

In other words, we only needed to consider the functions presented in the spacetime metric (\ref{metric}) into \eqref{potential} to obtain the short-ranged potential $V(b(b^{*}))$, given by \eqref{potentialbmm}, that will be considered in the next sections.

\subsection{The Quasinormal Modes of the massless spin 1/2 field for the BMM Black Hole}

In order to calculate the spectrum of complex frequencies, i.e. the quasinormal mode frequencies, of the Dirac field for the BMM Black Hole, we need to solve the master wave equation \eqref{masterequation} with the specific boundary conditions of purely outgoing waves at  infinity, and purely ingoing waves at the event horizon. So, the solution $\psi(b^{*})$ of Eq. \eqref{masterequation} must behave as:

\begin{align}
    \psi \sim e^{+i\omega b^{*}},\qquad b^{*}\rightarrow+\infty, \label{boundary1}\\
    \psi \sim e^{-i\omega b^{*}},\qquad b^{*}\rightarrow-\infty. \label{boundary2}
\end{align}

The set of complex frequencies $\omega$ which satisfy the equation \eqref{masterequation} and the boundary conditions above form the discrete spectrum of quasinormal frequencies.

To accomplish the calculation of the quasinormal frequencies of the BMM Black Hole, we employ the sixth order WKB method \cite{Konoplya_2003}. The WKB method, as proposed initially by Iyer and Will \cite{wkb1} up to third order, and posteriorly upgraded to sixth order by Konoplya \cite{Konoplya_2003}, is a semianalytic technique for determining the complex quasinormal mode frequencies of black holes for any kind of field perturbation, including gravitational ones. It already incorporates the boundary conditions of Eqs. \eqref{boundary1} and \eqref{boundary2}, and it actually manages to provide very accurate results \cite{Konoplya_2019} when compared to other calculation procedures.

The formula for the frequencies in the sixth order WKB method is the following:

\begin{equation}
    \frac{i(\omega^2-V(r_{max}))}{\sqrt{-2\partial^2_{*}V\rvert_{r_{max}}}}-\sum_{i=1}^{6}\Lambda_{i}=n+\frac{1}{2}, \label{wkb}
\end{equation}

\noindent where $n$ is the overtone number, $r_{max}$ is the point at which the potential reaches its maximum and the corrections $\Lambda_{i}$ contain the derivatives of the potential avaliated at $r_{max}$ up to the 12th order and are reported in \cite{Konoplya_2003,Konoplya_2019}. This formula is an improvement over the third order formula obtained in \cite{wkb1}. 

With equation \eqref{wkb}, we can finally evaluate the quasinormal frequencies of the massless Dirac field for the BMM Black Hole. For this purpose, we only need to replace the potential $V$ in that formula by one of the potentials derived in the previous section, and given by Eq. \eqref{potentialbmm}. We will choose this potential to be $V_{+}(b)$, without any loss of generality. An analog analysis can be made for $V_{-}(b)$, but since the behavior of $V_{-}(b)$ is qualitatively similar to $V_{+}(b)$, it is enough to consider $V_{+}(b)$. Accordingly, from now on:

\begin{equation}
    V(b)=
    V_{+}(b)=\frac{4}{\lambda_{2}^2}\left[\frac{\lambda_{1}\lambda_{2}M_{BH}}{2} \left(\frac{M_{BH}}{M_{WH}} \right)^{\frac{3}{2}}\right]^{\frac{1}{2}}\left[(l(l+1)+1/4)\frac{a(b)}{b^2}+ \sqrt{l(l+1)+1/4}\left(\frac{2}{\lambda_{2}}\right)\frac{a(b)}{\left(1+\frac{1}{X^2(b)}\right)\frac{dX}{db}}\left(\frac{\sqrt{a(b)}}{b} \right)'\right]. \label{potentialplusbmm}
\end{equation}

\begin{figure}[!ht]
    \centering
    \begin{subfigure}{0.5\textwidth}
    \centering
     \includegraphics[width=8cm]{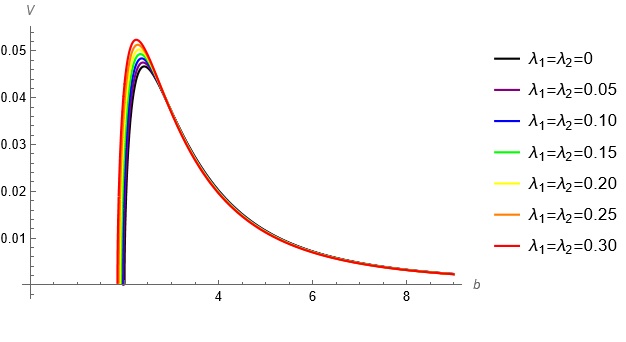}
    \caption{\centering $\beta=\frac{3}{5}$}
    \end{subfigure}%
     \begin{subfigure}{0.5\textwidth}
     \centering
     \includegraphics[width=8cm]{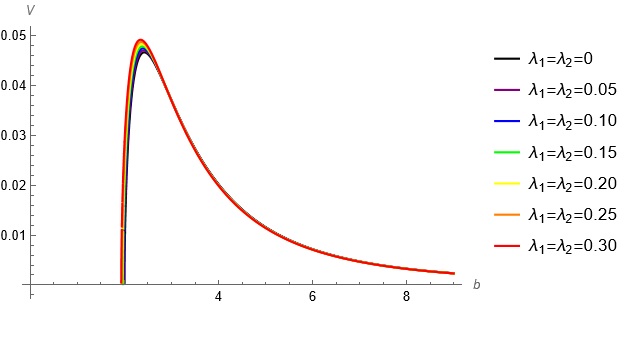}
    \caption{\centering $\beta=\frac{5}{3}$}
    \end{subfigure}
    \caption{Short-ranged potentials for massless Dirac fields for $l=\frac{1}{2}$}
    \label{fig:pot1}
\end{figure}

\begin{figure}[!ht]%
    \centering
    \begin{subfigure}{0.5\textwidth}
    \centering
    \includegraphics[width=8cm]{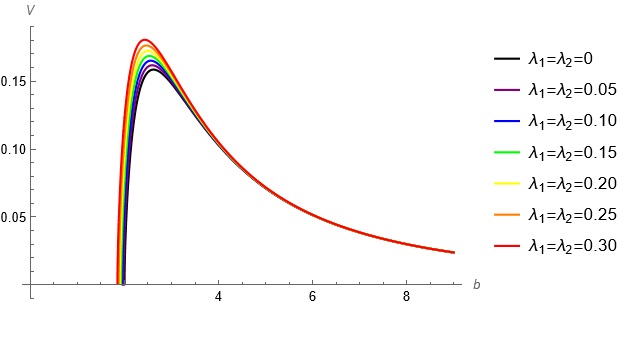}
    \caption{\centering $\beta=\frac{3}{5}$}
    \end{subfigure}%
    \begin{subfigure}{0.5\textwidth}
    \centering
    \includegraphics[width=8cm]{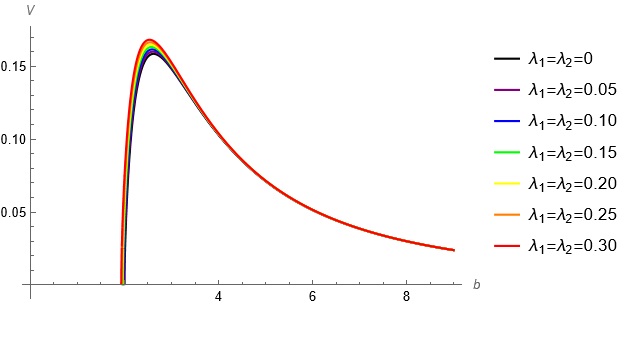}
    \caption{\centering $\beta=\frac{5}{3}$}
    \end{subfigure}%
    \caption{Short-ranged potentials for massless Dirac fields for $l=\frac{3}{2}$}%
    \label{fig:pot3}%
\end{figure}

\begin{figure}[!ht]%
    \centering
    \begin{subfigure}{0.5\textwidth}
    \centering
    \includegraphics[width=8cm]{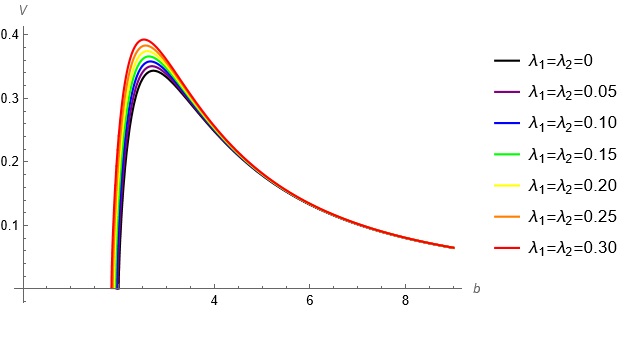}
    \caption{\centering $\beta=\frac{3}{5}$}
    \end{subfigure}%
    \begin{subfigure}{0.5\textwidth}
    \centering
    \includegraphics[width=8cm]{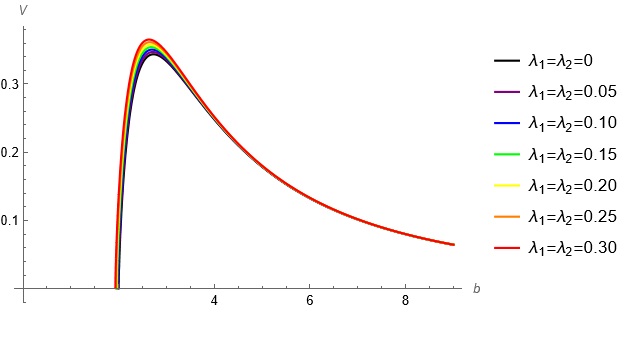}
    \caption{\centering $\beta=\frac{5}{3}$}
    \end{subfigure}%
    \caption{Short-ranged potentials for massless Dirac fields for $l=\frac{5}{2}$}%
    \label{fig:pot5}%
\end{figure}

This is the short-ranged potential that we will consider for our applications within this framework. In the following images we plotted some graphics of this potential for some reference values of $\lambda_{1}$ and $\lambda_{2}$ and always assuming the normalization $M_{BH}=1$. For simplicity, we chose $\lambda_{1}=\lambda_{2}=\lambda$ so we can analyse the behavior of the potential with a single varying parameter of LQG . For $\lambda=0$ we are led to the classical Schwarzschild case, which in those graphics will be represented by the black curves. As we can clearly see, as the LQG parameter $\lambda$ increases, the potential increases its maximum value. We assumed $l=1/2$ in Fig.\ref{fig:pot1}, $l=3/2$ in Fig.\ref{fig:pot3} and $l=5/2$ in Fig.\ref{fig:pot5}. These results are in accordance with the analysis of quasinormal modes of other fields, like the scalar, vector and tensor ones \cite{Bouhmadi-Lopez:2020oia}.

%\newpage

\subsection{Results and Graphics}

Now, we can finally go to the main discussion of the behaviour of the quasinormal frequencies of the massless Dirac field in a quantum corrected black hole spacetime inspired by the loop quantum gravity theory \cite{Bodendorfer_2019}. In order to obtain those frequencies, we applied the 6th order WKB method \eqref{wkb} to the potential given by Eq. \eqref{potentialplusbmm}. By doing that, we got that the quasinormal frequencies for the massless Dirac field surrounding the BMM black hole spacetime are given by the values listed in the tables shown in the Appendix at the end of this paper. Those values for the frequencies were evaluated for some reference values of the LQG parameters $\lambda_{1}$ and $\lambda_{2}$ and for a fixed value of $M_{BH}=1$. They are in good agreement with first analyses carried out in this context, like in \cite{Cho:2003qe}, with a $\sim 1\%$ difference due to the use of the third order approximation instead of our sixth order (their parameter $|\kappa|$ is related to our $l$ according to $\kappa^2=l(l+1)+1/4$).

\begin{figure}[!ht]%
    \centering
    \begin{subfigure}{0.5\textwidth}
    \centering
    \includegraphics[width=8cm]{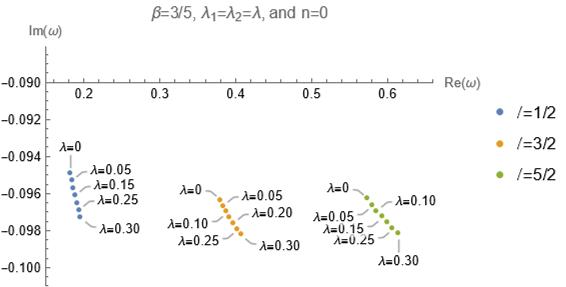}
    \caption{\centering $\beta=\frac{3}{5}$}%
    \end{subfigure}%
    \begin{subfigure}{0.5\textwidth}
    \centering
    \includegraphics[width=8cm]{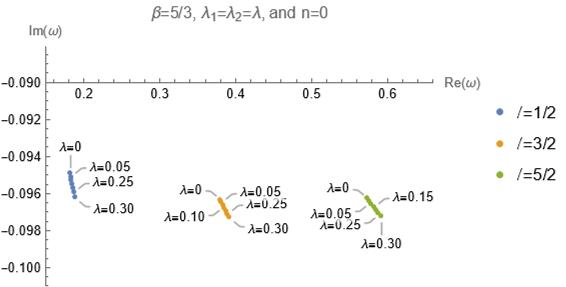}
    \caption{\centering $\beta=\frac{5}{3}$}
    \end{subfigure}%
    \caption{Complex frequency plane for $n=0$, with $\beta=\frac{3}{5}$ on the left and $\beta=\frac{5}{3}$ on the right.}%
    \label{fig:complex0}%
\end{figure}

\begin{figure}[!ht]%
    \centering
    \begin{subfigure}{0.5\textwidth}
    \centering
    \includegraphics[width=8cm]{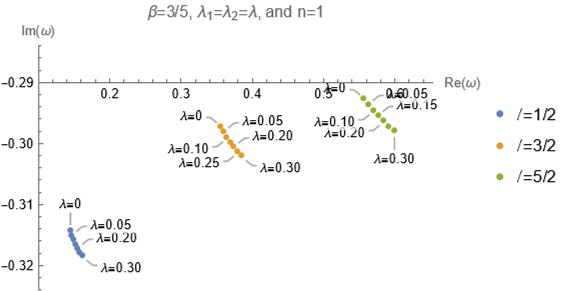}
    \caption{\centering $\beta=\frac{3}{5}$}
    \end{subfigure}%
    \begin{subfigure}{0.5\textwidth}
    \centering
    \includegraphics[width=8cm]{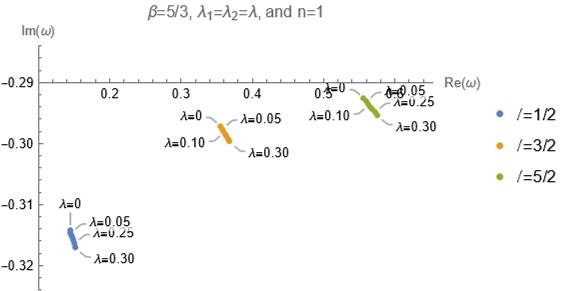}
    \caption{\centering $\beta=\frac{5}{3}$}
    \end{subfigure}%
    \caption{Complex frequency plane for $n=1$, with $\beta=\frac{3}{5}$ on the left and $\beta=\frac{5}{3}$ on the right.}%
    \label{fig:complex1}%
\end{figure}

\begin{figure}[!ht]%
    \centering
    \begin{subfigure}{0.5\textwidth}
    \centering
    \includegraphics[width=8cm]{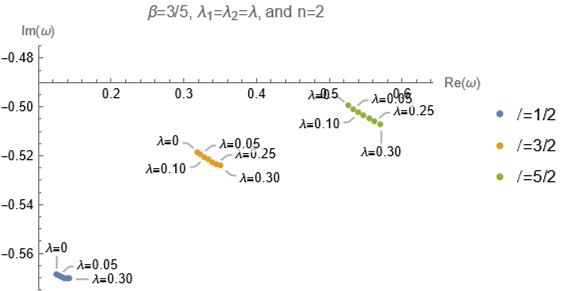}
    \caption{\centering $\beta=\frac{3}{5}$}%
    \end{subfigure}%
     \begin{subfigure}{0.5\textwidth}
     \centering
     \includegraphics[width=8cm]{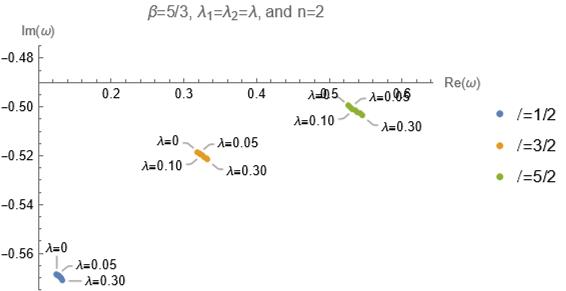}
    \caption{\centering $\beta=\frac{5}{3}$}
    \end{subfigure}%
    \caption{Complex frequency plane for $n=2$, with $\beta=\frac{3}{5}$ on the left and $\beta=\frac{5}{3}$ on the right.}%
    \label{fig:complex2}%
\end{figure}

We present at this section some graphics describing the behaviour of those quasinormal frequencies obtained here, as well as the physical discussion concerning their meanings. We can start by showing the behaviour of those frequencies in the complex frequency plane. In these graphics we show how the quasinormal frequencies for the massless Dirac field surrounding a static and spherically symmetric quantum corrected black hole deviate from their classical values given by the Schwarzschild metric (which here is obtained by the limit $\lambda_{1}=\lambda_{2}=0$ for the LQG parameters). For the sake of simplicity, we only show in the complex frequency plane, the quasinormal frequencies associated with symmetrical values of the LQG parameters, i.e. for $\lambda_{1}=\lambda_{2}=\lambda$. We took three reference values for $l$, i.e., $l=\frac{1}{2},\frac{3}{2},\frac{5}{2}$ for each graphic. In the following, from the first pair of graphics to the last one, we will distinguish among three different scenarios for the overtune numbers $n$. We consider the ground mode $n=0$ (Fig.(\ref{fig:complex0})), and the following two overtune numbers $n=1$ (Fig.(\ref{fig:complex1})) and $n=2$ (Fig.(\ref{fig:complex2})). The graphics on the left are always for $\beta=\frac{3}{5}$, while the graphics on the right are always for $\beta=\frac{5}{3}$.

We can clearly see here that the deviation from the Schwarzschild's reference value for the quasinormal frequency is enhanced as the value of the LQG parameter $\lambda$ increases. That is true both for $\beta=\frac{3}{5}$ and for $\beta=\frac{5}{3}$.

%\newpage

\begin{figure}[!ht]%
    \centering
     \begin{subfigure}{0.5\textwidth}
     \centering
     \includegraphics[width=8cm]{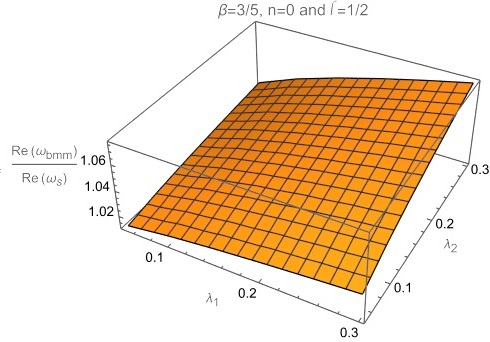}
    \caption{\centering $\beta=\frac{3}{5}$}
    \end{subfigure}%
    \begin{subfigure}{0.5\textwidth}
    \centering
    \includegraphics[width=8cm]{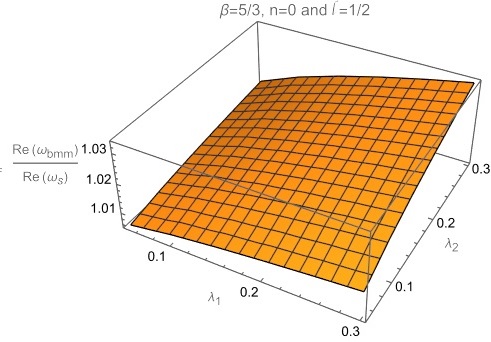}
    \caption{\centering $\beta=\frac{5}{3}$}
    \end{subfigure}%
    \caption{Ratio between the real parts of the QN frequencies of the BMM Black hole and the Schwarzschild's analog value. This graphic was plotted for $n=0$ and $l=\frac{1}{2}$. We have  $\beta=\frac{3}{5}$ on the left side and $\beta=\frac{5}{3}$ on the right.}%
    \label{fig:real}%
\end{figure}

\begin{figure}[!ht]%
    \centering
    \begin{subfigure}{0.5\textwidth}
    \centering
    \includegraphics[width=8cm]{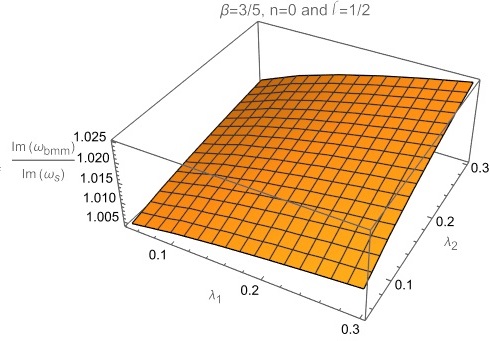}
    \caption{\centering $\beta=\frac{3}{5}$}
    \end{subfigure}%
    \begin{subfigure}{0.5\textwidth}
    \centering
    \includegraphics[width=8cm]{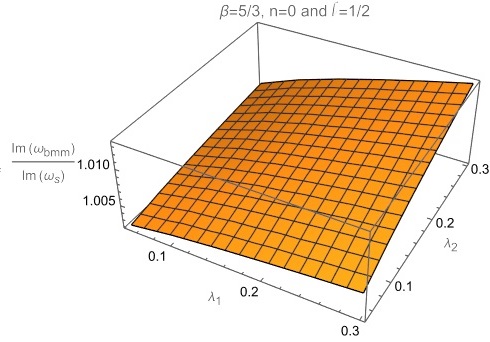}
    \caption{\centering $\beta=\frac{5}{3}$}
    \end{subfigure}%
    \caption{Ratio between the imaginary parts of the QN frequencies of the BMM Black hole and the Schwarzschild's analog value. This graphic was plotted for $n=0$ and $l=\frac{1}{2}$. We have  $\beta=\frac{3}{5}$ on the left side and $\beta=\frac{5}{3}$ on the right.}%
    \label{fig:imm}%
\end{figure}

For the ground mode specifically, we can also see how the quasinormal frequency values deviate from their Schwarzschild's reference value as we vary $\lambda_{1}$ and $\lambda_{2}$ at the same time. This is shown in the following three dimensional graphics bellow. As usually, in the left we have $\beta=\frac{3}{5}$ and in the right we have $\beta=\frac{5}{3}$. The first pair of graphics is the rate between the real part of the QN frequency for the BMM black hole and its respective Schwarzschild analog value (Fig.(\ref{fig:real})), while the second pair is going to be the imaginary part of the rate (Fig.(\ref{fig:imm})). We show above the rates between the real and imaginary parts of the ground mode of the QN frequency for the BMM case and the Schwarzschild case, with $l=\frac{1}{2}$. For $l=\frac{3}{2}$ and $\frac{5}{2}$, we find a similar behaviour, so we omit them here for simplicity.
%, while the second and third groups describe the same, but for $l=\frac{3}{2}$ and for $l=\frac{5}{2}$, respectively.

\begin{figure}[H]%
    \centering
    \begin{subfigure}{0.5\textwidth}
    \centering
    \includegraphics[width=8cm]{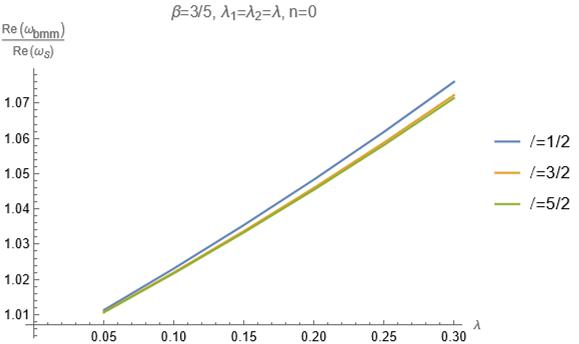}
    \caption{\centering $\beta=\frac{3}{5}$}
    \end{subfigure}%
    \begin{subfigure}{0.5\textwidth}
    \centering
    \includegraphics[width=8cm]{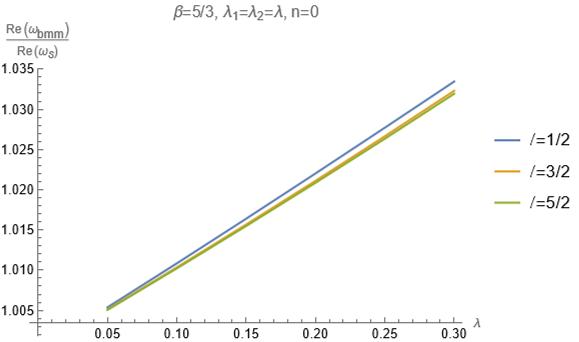}
    \caption{\centering $\beta=\frac{5}{3}$}
    \end{subfigure}%
    \caption{Ratio between the real parts of the QN frequencies of the BMM black hole and its Schwarzschild's analog value. This graphic was plotted by taking $\lambda_{1}=\lambda_{2}=\lambda$. For these graphics we have taken $n=0$ and $l=\frac{1}{2},\frac{3}{2},\frac{5}{2}$.}%
    \label{fig:real2}%
\end{figure}

\begin{figure}[H]%
    \centering
    \begin{subfigure}{0.5\textwidth}
    \centering
\includegraphics[width=8cm]{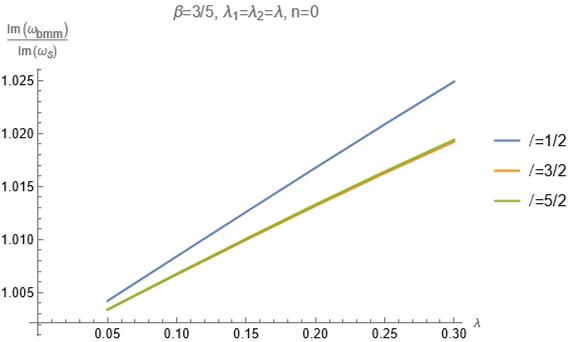}
    \caption{\centering $\beta=\frac{3}{5}$}
    \end{subfigure}%
     \begin{subfigure}{0.5\textwidth}
     \centering
     \includegraphics[width=8cm]{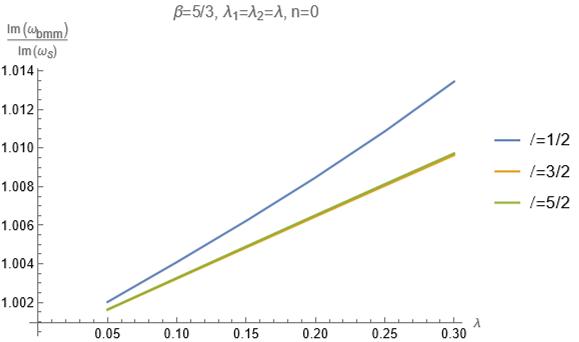}
     \caption{\centering $\beta=\frac{5}{3}$}
     \end{subfigure}%
    \caption{Ratio between the imaginary parts of the QN frequencies of the BMM black hole and its Schwarzschild's analog value. This graphic was plotted by taking $\lambda_{1}=\lambda_{2}=\lambda$. Here, we have also taken $n=0$ and $l=\frac{1}{2},\frac{3}{2},\frac{5}{2}$.}%
    \label{fig:imm2}%
\end{figure}

We can also plot 2-dimensional graphics out of the 3-dimensional ones if we take $\lambda_{1}=\lambda_{2}=\lambda$. This will allow us to compare the behaviour of the ratio between the quasinormal frequency for BMM Black Hole and Schwarzschild black hole for the three different values of $l$. Once again, we divide our analysis between $\beta=\frac{3}{5}$ and $\beta=\frac{5}{3}$, and for the real (Fig.(\ref{fig:real2})) and imaginary parts (Fig.(\ref{fig:imm2})) of the quasinormal modes.

From Figs.(\ref{fig:real2}) and (\ref{fig:imm2}), we see that the growth rate of the imaginary and real parts of $\omega$ with the LQG parameter basically does not distinguish between the cases $l=3/2$ and $l=5/2$. On the other hand, the lowest multipole number $l=1/2$ gives visibly more intense values of these parts and also grow at a higher rate when looking at the imaginary part of quasinormal frequency.

\begin{figure}[H]
\centering
\begin{subfigure}[b]{.49\linewidth}
\includegraphics[width=\linewidth]{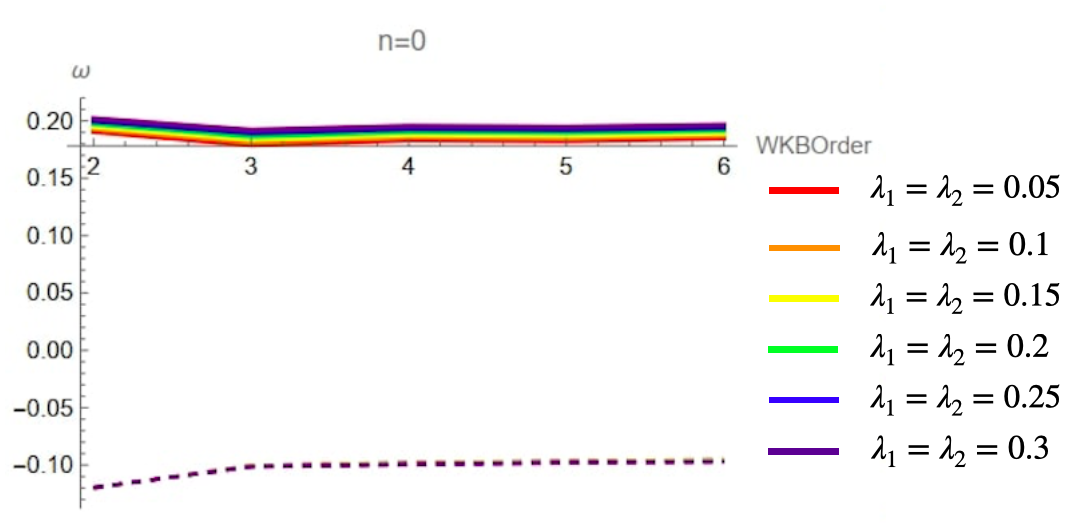}
\caption{$n=0$}\label{fig:stab-n0}
\end{subfigure}
\begin{subfigure}[b]{.49\linewidth}
\includegraphics[width=\linewidth]{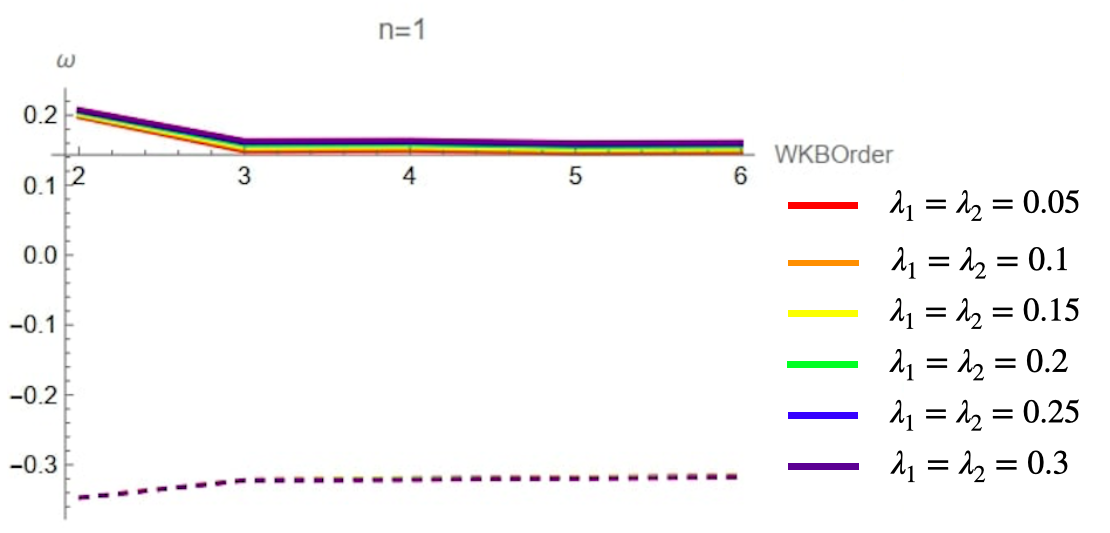}
\caption{$n=1$}\label{fig:stab-n1}
\end{subfigure}

\begin{subfigure}[b]{.49\linewidth}
\includegraphics[width=\linewidth]{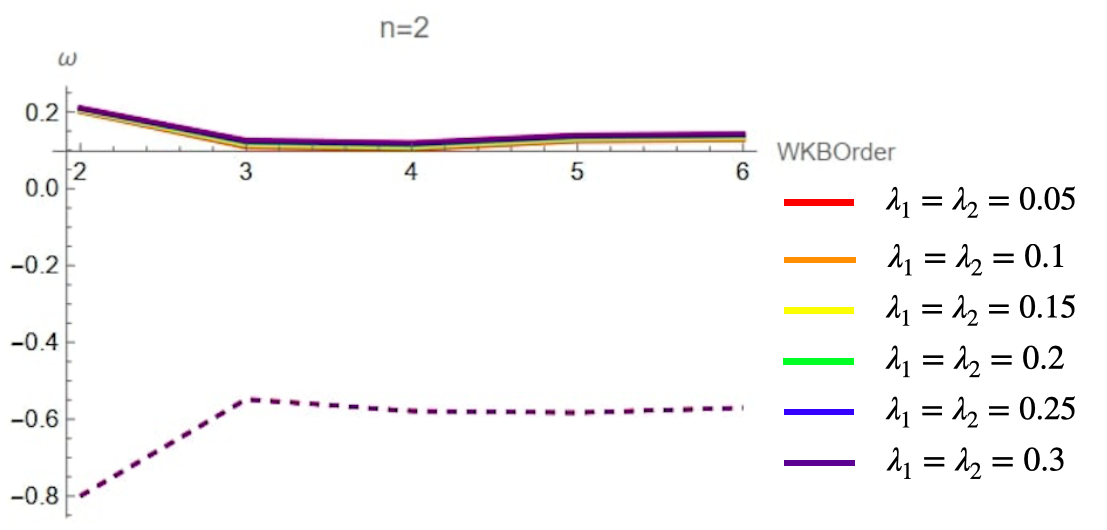}
\caption{$n=2$}\label{fig:stab-n2}
\end{subfigure}
\caption{Real (solid lines) and imaginary (dashed lines) parts of the quasinormal frequencies for different values of the quantum gravity parameters $\lambda_1$ and $\lambda_2$ versus the order of the WKB method. The imaginary parts are too close to each other and cannot be distinguished in this resolution.}
\label{fig:stab}
\end{figure}

It is interesting to test the validity of the WKB approach in this solution by analyzing the stability of quasinormal frequencies for low and high overtone numbers $n$. For this reason, in Figs.\ref{fig:stab-n0}, \ref{fig:stab-n1}, \ref{fig:stab-n2}, we show that the real and imaginary parts of the quasinormal frequencies tend to fixed values for different quantum gravity parameters $\lambda_1$ and $\lambda_2$ when the order of the WKB method grows (horizontal axis). We observe that despite the growth of $n$, the frequencies stabilize after the 3rd order and continues stable at the 6th one.

%\newpage
\section{The Hawking Radiation and the Greybody Factor}\label{sec:grey}

The equations describing the
perturbations of black hole spacetimes by test fields with spins 0 (scalar),1 (vectorial), 2
(tensorial), and 1/2 (spinorial) were extensively investigated \cite{Arbey:2021jif,Teukolsky,1973ApJ...185..649P,1974ApJ...193..443T,PhysRevD.13.198, PhysRevD.14.3260, PhysRevD.16.2402,1975RSPSA.345..145C,1976RSPSA.348...39C,1976RSPSA.350..165C,1977RSPSA.352..325C}. In these papers, the authors
derived the rates of emission of Hawking radiation for all those fields. For a Dirac test field in a
black hole spacetime, they found a Hawking radiation spectrum that obeys Fermi-Dirac
statistics; while for electromagnetic, scalar, and gravitational perturbations, they showed that the
Hawking radiation spectrum obeys exactly Bose-Einstein statistics.

A large literature has been produced to calculate those semiclassical properties of different
modified black holes spacetimes \cite{universe8010050}. The properties investigated by those works were
the variations, caused by the metric modifications, in the Hawking radiation spectrum of every
test field and in the temperature and entropy of the black holes. Having those results, the full
thermodynamics of those modified models of black holes were finally formulated, and the
signature of the metric modifications was then investigated in the final thermodynamical
description. Some of the most important works on this topic were the ones that considered
quantum corrections to black hole spacetime solutions.

Studying the physical consequences of Loop Quantum Gravity polymerised models on the
Hawking radiation and the thermodynamics of black holes requires us to derive and solve the
equations of motion of various perturbing fields on the background of LQG black hole
spacetimes. A previous paper was dedicated to study and describe the thermodynamics properties of the BMM black hole \cite{Mele:2021hro}. Another previous paper managed to describe the scalar, electromagnetic and gravitational radiation emitted by this black hole \cite{Bouhmadi-Lopez:2020oia}. Now, if we want to study the Dirac field radiation produced by a LQG black hole, we have to derive and solve the Dirac equation at the neighbourhood of the black hole described by this LQG model. Since this is the same problem that we faced in the previous section (when dealing with the quasinormal modes), part of the analytical development made before will be applied here again. The semi-analytical method used there could also be used here again. However, for the purpose of the qualitative discussion we plan to do here, another method seemed more promising. This method was the Bounding Bogoliubov Coefficients
method \cite{Visser_1999, Boonserm_2008,Shankaranarayanan_2003,Boonserm_2010}.

So, finally, in order to study the signature of Loop Quantum Gravity polymerised model on the radiation spectra and on the
Greybody Factor of the Dirac field of the LQG black hole, we need to recall the
equation of motion of this perturbing field on the background of our LQG black hole
spacetimes \cite{Bodendorfer_2019}. We derived this equation of motion in the previous section, and the final equation we found was \eqref{masterequation}, with the short-ranged potential given by \eqref{potentialbmm}.

%The bouding Bogoliubov Coefficients method claims that the actual Greybody Factor (or the transfer coefficient) of a test field surrounding a black hole should always be greater than or equal to the following expression for the short-ranged potential of this field:

The bouding Bogoliubov Coefficients method can be used to show that the actual Greybody Factor (or the transfer coefficient) of a test field surrounding a black hole should always be greater than or equal to a certain function of the potential \cite{Visser_1999,Boonserm_2008,Shankaranarayanan_2003,Boonserm_2010}. Relying on Ref.\cite{Visser_1999}, this can be found by expressing the solution of the Schr\"{o}dinger equation in terms of auxiliary functions $a(x)$, $b(x)$ and $\varphi(x)$
\begin{equation}
    \psi(x)=a(x)\frac{\exp(+i\varphi)}{\sqrt{\varphi'}}+b(x)\frac{\exp(-i\varphi)}{\sqrt{\varphi'}}\, ,
\end{equation}
where the values of the functions $a$ and $b$ at the boundary of the spacetime (i.e., the event horizon and region asymptotic far from the black hole) give the Bogoliubov coefficients $\alpha$ and $\beta$, respectively. The transmission probability is then given by $\sigma_l=|\alpha|^{-2}$.

The method is carried on by expressing the the second order Schr\"{o}dinger equation equivalently in terms of two first order differential equations for $a$ and $b$. The equation for $a$ can be simplified by the use of the Cauchy-Schwarz inequality (namely $|\text{Re}(A\, B)|\leq |A\, B|\leq |A|\, |B|$ $\forall\, A,\, B\in {\mathbb C}$) to become the inequality
\begin{equation}
    \frac{d|a|}{dx}\leq\varsigma \sqrt{|a|^2-1}\, .
\end{equation}
where 
\begin{equation}
    \varsigma=\frac{\sqrt{(h')^2+(\omega^2-V-h^2)^2}}{2h}\,,
\end{equation}
and we have defined the positive function $h\doteq\varphi'$. This inequality is the origin of the bound on the Greybody Factor. An integration of this expression at the boundary of the spacetime gives an inequality for $|\alpha|^{-1}=\sqrt{\sigma_l}$, which becomes a bound on the transmission probability
\begin{equation}
    \sigma_{l}\geq \sech^2\left(\int_{-\infty}^{\infty} \varsigma dr_{*} \right).
\end{equation}

When we set $h=\omega$ (the transmission coefficient), then
\begin{equation}
    \sigma_{l}\geq \sech^2\left(\frac{1}{2\omega}\int_{-\infty}^{\infty} V dr_{*} \right).
\end{equation}

\begin{figure}[!ht]%
    \centering
\begin{subfigure}{0.5\textwidth}
\centering
\includegraphics[width=8cm]{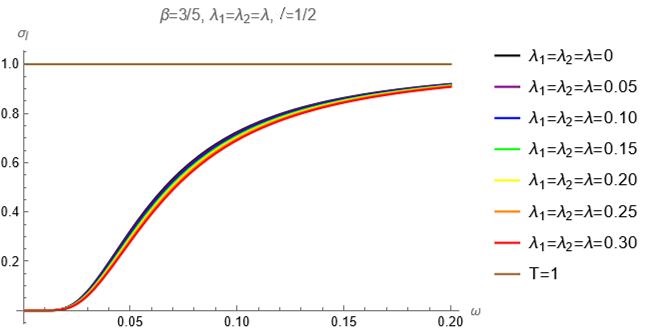}
    \caption{\centering $\beta=\frac{3}{5}$}
    \end{subfigure}%
    \begin{subfigure}{0.5\textwidth}
\centering
\includegraphics[width=8cm]{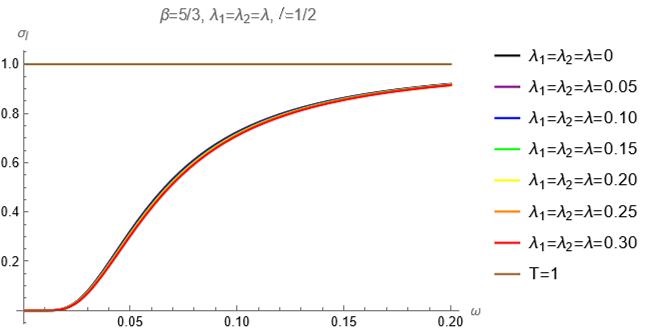}
    \caption{\centering $\beta=\frac{5}{3}$}
    \end{subfigure}%
    \caption{Minimum value for the Greybody Factor (transfer coefficient) for the massless Dirac radiation for the BMM black hole. We have taken $\lambda_{1}=\lambda_{2}=\lambda$, and $l=\frac{1}{2}$. For the left graphic we have $\beta=\frac{3}{5}$, and for the right one, we have $\beta=\frac{5}{3}$. The brown line is basically the ceiling-value of $1$. }%
    \label{fig:grey1}%
\end{figure}

\begin{figure}[!ht]%
    \centering
     \begin{subfigure}{0.5\textwidth}
     \centering
     \includegraphics[width=8cm]{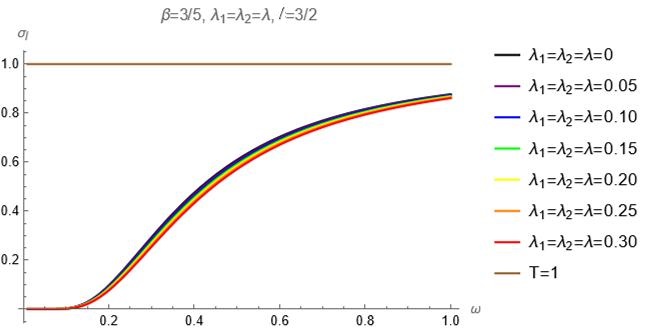}
    \caption{\centering $\beta=\frac{3}{5}$}%
    \end{subfigure}%
     \begin{subfigure}{0.5\textwidth}
     \centering
     \includegraphics[width=8cm]{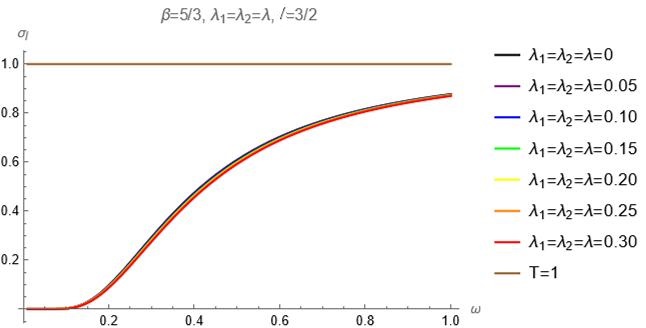} 
    \caption{\centering $\beta=\frac{5}{3}$}
    \end{subfigure}%
    \caption{Minimum value for the Greybody Factor (transfer coefficient) for the massless Dirac radiation for the BMM black hole. We have taken $\lambda_{1}=\lambda_{2}=\lambda$, and $l=\frac{3}{2}$. For the left graphic we have $\beta=\frac{3}{5}$, and for the right one, we have $\beta=\frac{5}{3}$. The  brown line is basically the ceiling-value of $1$. }%
    \label{fig:grey3}%
\end{figure}

\begin{figure}[!ht]%
    \centering
     \begin{subfigure}{0.5\textwidth}
     \centering
     \includegraphics[width=8cm]{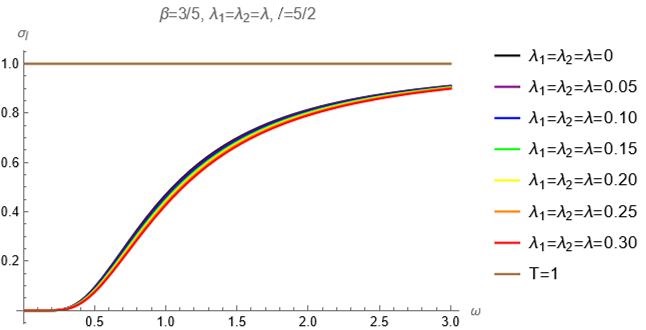}
    \caption{\centering $\beta=\frac{3}{5}$}
    \end{subfigure}%
    \begin{subfigure}{0.5\textwidth}
     \centering
     \includegraphics[width=8cm]{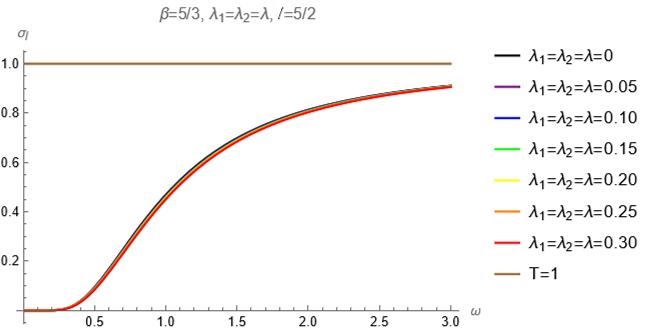}
    \caption{\centering $\beta=\frac{5}{3}$}
    \end{subfigure}%
    \caption{Minimum value for the Greybody Factor (transfer coefficient) for the massless Dirac radiation for the BMM black hole. We have taken $\lambda_{1}=\lambda_{2}=\lambda$, and $l=\frac{5}{2}$. For the left graphic we have $\beta=\frac{3}{5}$, and for the right one, we have $\beta=\frac{5}{3}$. The brown line is basically the ceiling-value of $1$. }%
    \label{fig:grey5}%
\end{figure}

Hence, substituting $V$ in the expression above by \eqref{potentialbmm}, we can obtain that the GF for the massless Dirac radiation for the BMM black hole should always be greater than or equal to the following curves plotted in Figs.(\ref{fig:grey1}), (\ref{fig:grey3}) and (\ref{fig:grey5}).

%%%%%%%%%%%%%%%%%%%%%%%%%%%%%%%%%%%%%%%%%%%%%%%%%%%%%%%%%%%%%%%%%%%%%%%%%%%%%%%%%%%%%%%%%%%%%%%%

%%%%%%%%%%%%%%%%%%%%%%%%%%%%%%%%%%%%%%%%%%%%%%%%%%%%%%%%%%%%%%%%%%%%%%%%%%%%%%%%%%%%%%%%%%%%%%%%

%%%%%%%%%%%%%%%%%%%%%%%%%%%%%%%%%%%%%%%%%%%%%%%%%%%%%%%%%%%%%%%%%%%%%%%%%%%%%%%%%%%%%%%%%%%%%%%%

%%%%%%%%%%%%%%%%%%%%%%%%%%%%%%%%%%%%%%%%%%%%%%%%%%%%%%%%%%%%%%%%%%%%%%%%%%%%%%%%%%%%%%%%%%%%%%%%
%\newpage

As we could see from those graphics, an increase in the value of the LQG parameter $\lambda$ implied in a decrease of the minimum value of the Greybody Factor (or transfer rate) of the massless Dirac field as seen by an assymptotic observer. Accordingly, increasing the value of the LQG parameter implies in a Dirac radiation spectrum, emitted by the BMM black hole and measured by a distant observer, even more distant to the the ideal Blackbody radiation spectrum. In all graphics, we compared our results with the Schwarzschild case ($\lambda=0$) and with the ideal black body radiation $(T=1)$. The reduction of the Greybody Factor means that quantum gravity contributions are facilitating the absorption of particles by the black hole.

\section{Final Remarks}\label{sec:final}

Finally, we see that in this framework, we managed to generalize the previous analysis \cite{Bouhmadi-Lopez:2020oia} of the quasinormal modes of the masslass scalar field, electromagnetic field and the gravitational field, in the framework of a consistent model of non-singular Schwarzschild black hole in LQG \cite{bodendorfer2019effective}, by studying the quasinormal modes of the massless Dirac field as well. The calculations performed were made through the 6th order WKB method. We have determined the impact of the LQG parameters on the potential and on the real and imaginary parts of the quasinormal frequencies of a black hole with mass parameter $M_{BH}=1$, for some values of the overtune numbers $n$ ranging from $0$ to $2$, for the multipole number $l$ ranging from $1/2$ to $5/2$ and for the two possibles cases that characterize such black hole regarding ``mass amplification'' ($\beta=5/3$) and ``mass de-amplification'' ($\beta=3/5$). Overall, we have found an increment on the potential and on the real and imaginary parts of the quasinormal frequencies when the LQG parameters grow.

Furthermore, as another application of the master wave equation \eqref{masterequation}, we managed to study and discuss the Greybody Factor of the Hawking radiation associated with the massless Dirac field surrounding this black hole. This last application was made by means of bounding the Bogoliubov coefficients. We have found a global reduction on the lower bound of the greybody factor with the growth of the LQG parameters, which is more prominent for the mass de-amplification case, i.e., for the case in which one has a reduction of the mass of the white hole spacetime perceived by an observer who transits from the black hole to the white hole spacetime.

As discussed in \cite{Bouhmadi-Lopez:2020oia}, since the parameters $\lambda_1$ and $\lambda_2$ are such that $\lambda_1\lambda_2$ are of mass dimension $3$, the dimensionless factor that governs modifications of the equations are of the order $\lambda_1\lambda_2/M_{BH}^3\sim M_{\text{Planck}}^3/M_{BH}$, which for a solar mass black hole would be very small. Nevertheless, for Planckian black holes, the deviations from general relativity predictions would be of the order of $7\%$ or higher, as can be seen from Fig.(8). It would be interesting to further investigate such black hole proposals in order to seek for phenomena that could somehow amplify these third order Planck-scale effects even for third order corrections, which has been object of investigation through the presence of amplifiers in different contexts \cite{PierreAuger:2021mve,Lobo:2021yem,Lobo:2020qoa}.

%%%%%%%%%%%%%%%%%%%%%%%%%%%%%%%%%%%%%%%%%%%%%%%%%%%%%%%%%%%%%%%%%%%%%%%%%%%%%%%%%%%%%%%%%%%%%%%%

\section*{Acknowledgements}
S.A. thank Coordena\c c\~ao de Aperfei\c coamento de Pessoal de N\'ivel Superior - Brazil (CAPES) - Finance Code 001 for financial support. S. A. would like to acknowledge the contribution of Roman Konoplya for the disposal of the code for the calculations. I. P. L. was partially supported by the National Council for Scientific and Technological Development - CNPq grant 306414/2020-1 and by the grant 3197/2021, Para\'iba State Research Foundation (FAPESQ). I. P. L. would like to acknowledge the contribution of the COST Action CA18108. V.B.B. is partially supported by CNPq through the Research Project No. 307211/2020-7.
%%%%%%%%%%%%%%%%%%%%%%%%%%%%%%%%%%%%%%%%%%%%%%%%%%%%%%%%%%%%%%%%%%%%%%%%%%%%%%%%%%%%%%%%%%%%%%%%

\bibliographystyle{utphys}
\bibliography{quasinormal}

\providecommand{\href}[2]{#2}\begingroup\raggedright\begin{thebibliography}{}

\bibitem{Rova}
Rovelli, Carlo, ``{Loop quantum gravity},''
  \href{http://dx.doi.org/10.12942/lrr-1998-1}{{\em Living Rev. Rel.}
  {\bfseries 1} (1998) 1}, \href{http://arxiv.org/abs/gr-qc/9710008}{{\ttfamily
  arXiv:gr-qc/9710008}}.

\bibitem{Ashtekar_2004}
Ashtekar, Abhay and Lewandowski, Jerzy, ``{Background independent quantum
  gravity: A Status report},''
  \href{http://dx.doi.org/10.1088/0264-9381/21/15/R01}{{\em Class. Quant.
  Grav.} {\bfseries 21} (2004) R53},
  \href{http://arxiv.org/abs/gr-qc/0404018}{{\ttfamily arXiv:gr-qc/0404018}}.

\bibitem{https://doi.org/10.48550/arxiv.hep-th/0303185}
Smolin, Lee, ``{How far are we from the quantum theory of gravity?},''
  \href{http://arxiv.org/abs/hep-th/0303185}{{\ttfamily arXiv:hep-th/0303185}}.

\bibitem{Thiemann_2003}
Thiemann, Thomas, ``{Lectures on loop quantum gravity},''
  \href{http://dx.doi.org/10.1007/978-3-540-45230-0_3}{{\em Lect. Notes Phys.}
  {\bfseries 631} (2003) 41--135},
  \href{http://arxiv.org/abs/gr-qc/0210094}{{\ttfamily arXiv:gr-qc/0210094}}.

\bibitem{Ashtekar_2013}
Abhay Ashtekar,
  \href{http://dx.doi.org/10.1007/978-3-642-33036-0_2}{``Introduction to loop
  quantum gravity and cosmology,''} in {\em Quantum Gravity and Quantum
  Cosmology}, pp.~31--56.
\newblock Springer Berlin Heidelberg, 2013.
\newblock \url{https://doi.org/10.1007%2F978-3-642-33036-0_2}.

\bibitem{Gambini:2011zz}
Rodolfo Gambini, Jorge Pullin, {\em A First Course in Loop Quantum Gravity}.
\newblock Oxford University Press, 1~ed., 2011.

\bibitem{Rovelli_1995}
Carlo Rovelli and Lee Smolin, ``Discreteness of area and volume in quantum
  gravity,'' \href{http://dx.doi.org/10.1016/0550-3213(95)00150-q}{{\em Nuclear
  Physics B} {\bfseries 442} no.~3, (May, 1995) 593--619}.
  \url{https://doi.org/10.1016%2F0550-3213%2895%2900150-q}.

\bibitem{camelia}
Amelino-Camelia, Giovanni, ``Quantum-spacetime phenomenology,''
  \href{http://dx.doi.org/10.12942/lrr-2013-5}{{\em Living Reviews in
  Relativity} {\bfseries 16} (06, 2013) 5--}.

\bibitem{Addazi:2021xuf}
Addazi, A. and others, ``{Quantum gravity phenomenology at the dawn of the
  multi-messenger era\textemdash{}A review},''
  \href{http://dx.doi.org/10.1016/j.ppnp.2022.103948}{{\em Prog. Part. Nucl.
  Phys.} {\bfseries 125} (2022) 103948},
  \href{http://arxiv.org/abs/2111.05659}{{\ttfamily arXiv:2111.05659
  [hep-ph]}}.

\bibitem{Modesto_2004}
Leonardo Modesto, ``Disappearance of the black hole singularity in loop quantum
  gravity,'' \href{http://dx.doi.org/10.1103/physrevd.70.124009}{{\em Physical
  Review D} {\bfseries 70} no.~12, (Dec, 2004) }.
  \url{https://doi.org/10.1103%2Fphysrevd.70.124009}.

\bibitem{Modesto_2010}
Leonardo Modesto, ``Semiclassical loop quantum black hole,''
  \href{http://dx.doi.org/10.1007/s10773-010-0346-x}{{\em International Journal
  of Theoretical Physics} {\bfseries 49} no.~8, (Apr, 2010) 1649--1683}.
  \url{https://doi.org/10.1007%2Fs10773-010-0346-x}.

\bibitem{modesto2008black}
Modesto, Leonardo, ``{Black hole interior from loop quantum gravity},''
  \href{http://dx.doi.org/10.1155/2008/459290}{{\em Adv. High Energy Phys.}
  {\bfseries 2008} (2008) 459290},
  \href{http://arxiv.org/abs/gr-qc/0611043}{{\ttfamily arXiv:gr-qc/0611043}}.

\bibitem{bodendorfer2019effective}
Bodendorfer, Norbert and Mele, Fabio M. and M\"unch, Johannes, ``{Effective
  Quantum Extended Spacetime of Polymer Schwarzschild Black Hole},''
  \href{http://dx.doi.org/10.1088/1361-6382/ab3f16}{{\em Class. Quant. Grav.}
  {\bfseries 36} no.~19, (2019) 195015},
  \href{http://arxiv.org/abs/1902.04542}{{\ttfamily arXiv:1902.04542 [gr-qc]}}.

\bibitem{ashtekar1995quantization}
Ashtekar, Abhay and Lewandowski, Jerzy and Marolf, Donald and Mourao, Jose and
  Thiemann, Thomas, ``Quantization of diffeomorphism invariant theories of
  connections with local degrees of freedom,'' {\em Journal of Mathematical
  Physics} {\bfseries 36} no.~11, (1995) 6456--6493.

\bibitem{ashtekar2003mathematical}
Ashtekar, Abhay and Bojowald, Martin and Lewandowski, Jerzy, ``{Mathematical
  structure of loop quantum cosmology},''
  \href{http://dx.doi.org/10.4310/ATMP.2003.v7.n2.a2}{{\em Adv. Theor. Math.
  Phys.} {\bfseries 7} no.~2, (2003) 233--268},
  \href{http://arxiv.org/abs/gr-qc/0304074}{{\ttfamily arXiv:gr-qc/0304074}}.

\bibitem{ashtekar1997quantum}
Ashtekar, Abhay and Lewandowski, Jerzy, ``{Quantum theory of geometry. 1: Area
  operators},'' \href{http://dx.doi.org/10.1088/0264-9381/14/1A/006}{{\em
  Class. Quant. Grav.} {\bfseries 14} (1997) A55--A82},
  \href{http://arxiv.org/abs/gr-qc/9602046}{{\ttfamily arXiv:gr-qc/9602046}}.

\bibitem{cardosothesis}
Cardoso, Vitor, ``Quasinormal modes and gravitational radiation in black hole
  spacetimes,'' 2004.
\newblock \url{https://arxiv.org/abs/gr-qc/0404093}.

\bibitem{Vishveshwara:1970zz}
Vishveshwara, C. V., ``{Scattering of Gravitational Radiation by a
  Schwarzschild Black-hole},'' \href{http://dx.doi.org/10.1038/227936a0}{{\em
  Nature} {\bfseries 227} (1970) 936--938}.

\bibitem{cardoso2}
Davis, M. and Ruffini, R. and Press, W. H. and Price, R. H., ``{Gravitational
  radiation from a particle falling radially into a Schwarzschild black
  hole},'' \href{http://dx.doi.org/10.1103/PhysRevLett.27.1466}{{\em Phys. Rev.
  Lett.} {\bfseries 27} (1971) 1466--1469}.

\bibitem{cardoso3}
Gleiser, Reinaldo J. and Nicasio, Carlos O. and Price, Richard H. and Pullin,
  Jorge, ``{Colliding black holes: How far can the close approximation go?},''
  \href{http://dx.doi.org/10.1103/PhysRevLett.77.4483}{{\em Phys. Rev. Lett.}
  {\bfseries 77} (1996) 4483--4486},
  \href{http://arxiv.org/abs/gr-qc/9609022}{{\ttfamily arXiv:gr-qc/9609022}}.

\bibitem{cardoso4}
Anninos, Peter and Hobill, David and Seidel, Edward and Smarr, Larry and Suen,
  Wai-Mo, ``Collision of two black holes,''
  \href{http://dx.doi.org/10.1103/PhysRevLett.71.2851}{{\em Phys. Rev. Lett.}
  {\bfseries 71} (Nov, 1993) 2851--2854}.
  \url{https://link.aps.org/doi/10.1103/PhysRevLett.71.2851}.

\bibitem{Gan:2020dkb}
Gan, Wen-Cong and Santos, Nilton O. and Shu, Fu-Wen and Wang, Anzhong,
  ``{Properties of the spherically symmetric polymer black holes},''
  \href{http://dx.doi.org/10.1103/PhysRevD.102.124030}{{\em Phys. Rev. D}
  {\bfseries 102} (2020) 124030},
  \href{http://arxiv.org/abs/2008.09664}{{\ttfamily arXiv:2008.09664 [gr-qc]}}.

\bibitem{Bouhmadi-Lopez:2020oia}
Bouhmadi-L\'opez, Mariam and Brahma, Suddhasattwa and Chen, Che-Yu and Chen,
  Pisin and Yeom, Dong-han, ``{A consistent model of non-singular Schwarzschild
  black hole in loop quantum gravity and its quasinormal modes},''
  \href{http://dx.doi.org/10.1088/1475-7516/2020/07/066}{{\em JCAP} {\bfseries
  07} (2020) 066}, \href{http://arxiv.org/abs/2004.13061}{{\ttfamily
  arXiv:2004.13061 [gr-qc]}}.

\bibitem{Chen_2011}
Ju-Hua Chen and Yong-Jiu Wang, ``Complex frequencies of a massless scalar field
  in loop quantum black hole spacetime,''
  \href{http://dx.doi.org/10.1088/1674-1056/20/3/030401}{{\em Chinese Physics
  B} {\bfseries 20} no.~3, (Mar, 2011) 030401}.
  \url{https://doi.org/10.1088/1674-1056/20/3/030401}.

\bibitem{Santos}
Santos, Victor and Maluf, R. V. and Almeida, C. A. S., ``Quasinormal
  frequencies of self-dual black holes,''
  \href{http://dx.doi.org/10.1103/PhysRevD.93.084047}{{\em Phys. Rev. D}
  {\bfseries 93} (Apr, 2016) 084047}.
  \url{https://link.aps.org/doi/10.1103/PhysRevD.93.084047}.

\bibitem{Cruz:2020emz}
Cruz, M. B. and Brito, F. A. and Silva, C. A. S., ``{Polar gravitational
  perturbations and quasinormal modes of a loop quantum gravity black hole},''
  \href{http://dx.doi.org/10.1103/PhysRevD.102.044063}{{\em Phys. Rev. D}
  {\bfseries 102} no.~4, (2020) 044063},
  \href{http://arxiv.org/abs/2005.02208}{{\ttfamily arXiv:2005.02208 [gr-qc]}}.

\bibitem{qnmsrotatinglqg}
Liu, Cheng and Zhu, Tao and Wu, Qiang and Jusufi, Kimet and Jamil, Mubasher and
  Azreg-A\"{\i}nou, Mustapha and Wang, Anzhong, ``Shadow and quasinormal modes
  of a rotating loop quantum black hole,''
  \href{http://dx.doi.org/10.1103/PhysRevD.101.084001}{{\em Phys. Rev. D}
  {\bfseries 101} (Apr, 2020) 084001}.
  \url{https://link.aps.org/doi/10.1103/PhysRevD.101.084001}.

\bibitem{Konoplya_2003}
R. A. Konoplya, ``Quasinormal behavior of the $d$-dimensional schwarzschild
  black hole and the higher order {WKB} approach,''
  \href{http://dx.doi.org/10.1103/physrevd.68.024018}{{\em Physical Review D}
  {\bfseries 68} no.~2, (Jul, 2003) }.
  \url{https://doi.org/10.1103%2Fphysrevd.68.024018}.

\bibitem{Visser_1999}
Matt Visser, ``Some general bounds for one-dimensional scattering,''
  \href{http://dx.doi.org/10.1103/physreva.59.427}{{\em Physical Review A}
  {\bfseries 59} no.~1, (Jan, 1999) 427--438}.
  \url{https://doi.org/10.1103%2Fphysreva.59.427}.

\bibitem{Boonserm_2008}
Petarpa Boonserm and Matt Visser, ``Bounding the bogoliubov coefficients,''
  \href{http://dx.doi.org/10.1016/j.aop.2008.02.002}{{\em Annals of Physics}
  {\bfseries 323} no.~11, (Nov, 2008) 2779--2798}.
  \url{https://doi.org/10.1016%2Fj.aop.2008.02.002}.

\bibitem{Shankaranarayanan_2003}
S. Shankaranarayanan, ``Temperature and entropy of schwarzschild{\textendash}de
  sitter space-time,'' \href{http://dx.doi.org/10.1103/physrevd.67.084026}{{\em
  Physical Review D} {\bfseries 67} no.~8, (Apr, 2003) }.
  \url{https://doi.org/10.1103%2Fphysrevd.67.084026}.

\bibitem{Boonserm_2010}
Petarpa Boonserm and Matt Visser, ``Analytic bounds on transmission
  probabilities,'' \href{http://dx.doi.org/10.1016/j.aop.2010.02.005}{{\em
  Annals of Physics} {\bfseries 325} no.~7, (Jul, 2010) 1328--1339}.
  \url{https://doi.org/10.1016%2Fj.aop.2010.02.005}.

\bibitem{Bodendorfer_2019}
Norbert Bodendorfer and Fabio M Mele and Johannes MÃŒnch, ``A note on the
  hamiltonian as a polymerisation parameter,''
  \href{http://dx.doi.org/10.1088/1361-6382/ab32ba}{{\em Classical and Quantum
  Gravity} {\bfseries 36} no.~18, (Aug, 2019) 187001}.
  \url{https://doi.org/10.1088/1361-6382/ab32ba}.

\bibitem{Ashtekar_2011}
Abhay Ashtekar and Parampreet Singh, ``Loop quantum cosmology: a status
  report,'' \href{http://dx.doi.org/10.1088/0264-9381/28/21/213001}{{\em
  Classical and Quantum Gravity} {\bfseries 28} no.~21, (Sep, 2011) 213001}.
  \url{https://doi.org/10.1088%2F0264-9381%2F28%2F21%2F213001}.

\bibitem{oriti2017bouncing}
Oriti, Daniele and Sindoni, Lorenzo and Wilson-Ewing, Edward, ``{Bouncing
  cosmologies from quantum gravity condensates},''
  \href{http://dx.doi.org/10.1088/1361-6382/aa549a}{{\em Class. Quant. Grav.}
  {\bfseries 34} no.~4, (2017) 04LT01},
  \href{http://arxiv.org/abs/1602.08271}{{\ttfamily arXiv:1602.08271 [gr-qc]}}.

\bibitem{ashtekar2017loop}
Ashtekar, Abhay and Pullin, Jorge, {\em Loop Quantum Gravity: the first 30
  years}, vol.~4.
\newblock World Scientific, 2017.

\bibitem{ashtekar2006quantum}
Ashtekar, Abhay and Pawlowski, Tomasz and Singh, Parampreet, ``{Quantum Nature
  of the Big Bang: Improved dynamics},''
  \href{http://dx.doi.org/10.1103/PhysRevD.74.084003}{{\em Phys. Rev. D}
  {\bfseries 74} (2006) 084003},
  \href{http://arxiv.org/abs/gr-qc/0607039}{{\ttfamily arXiv:gr-qc/0607039}}.

\bibitem{ashtekar2015loop}
Ashtekar, Abhay and Barrau, Aurelien, ``{Loop quantum cosmology: From
  pre-inflationary dynamics to observations},''
  \href{http://dx.doi.org/10.1088/0264-9381/32/23/234001}{{\em Class. Quant.
  Grav.} {\bfseries 32} no.~23, (2015) 234001},
  \href{http://arxiv.org/abs/1504.07559}{{\ttfamily arXiv:1504.07559 [gr-qc]}}.

\bibitem{constraintsselfdualblackhole}
Yan, Jian-Ming and Wu, Qiang and Liu, Cheng and Zhu, Tao and Wang, Anzhong,
  ``Constraints on self-dual black hole in loop quantum gravity with s0-2 star
  in the galactic center,'' 2022.
\newblock \url{https://arxiv.org/abs/2203.03203}.

\bibitem{rovelli1990loop}
Rovelli, Carlo and Smolin, Lee, ``Loop space representation of quantum general
  relativity,'' {\em Nuclear Physics B} {\bfseries 331} no.~1, (1990) 80--152.

\bibitem{Caravelli_2010}
Francesco Caravelli and Leonardo Modesto, ``Spinning loop black holes,''
  \href{http://dx.doi.org/10.1088/0264-9381/27/24/245022}{{\em Classical and
  Quantum Gravity} {\bfseries 27} no.~24, (Nov, 2010) 245022}.
  \url{https://doi.org/10.1088%2F0264-9381%2F27%2F24%2F245022}.

\bibitem{Brahma_2018}
Suddhasattwa Brahma and Dong-Han Yeom, ``Effective black-to-white hole bounces:
  the cost of surgery,'' \href{http://dx.doi.org/10.1088/1361-6382/aae1df}{{\em
  Classical and Quantum Gravity} {\bfseries 35} no.~20, (Oct, 2018) 205007}.
  \url{https://doi.org/10.1088/1361-6382/aae1df}.

\bibitem{bojowaldrage}
Bojowald, Martin, ``{Comment (2) on ''Quantum Transfiguration of Kruskal Black
  Holes''},'' \href{http://arxiv.org/abs/1906.04650}{{\ttfamily
  arXiv:1906.04650 [gr-qc]}}.

\bibitem{towconsbtwhbfmc}
Achour, Jibril Ben and Brahma, Suddhasattwa and Mukohyama, Shinji and Uzan,
  Jean-Philippe, ``Towards consistent black-to-white hole bounces from matter
  collapse,''. \url{https://arxiv.org/abs/2004.12977}.

\bibitem{Achour_2020}
Jibril Ben Achour and Suddhasattwa Brahma and Jean-Philippe Uzan, ``Bouncing
  compact objects. part i. quantum extension of the oppenheimer-snyder
  collapse,'' \href{http://dx.doi.org/10.1088/1475-7516/2020/03/041}{{\em
  Journal of Cosmology and Astroparticle Physics} {\bfseries 2020} no.~03,
  (Mar, 2020) 041--041}. \url{https://doi.org/10.1088/1475-7516/2020/03/041}.

\bibitem{Ben_Achour_2020}
Jibril Ben Achour and Jean-Philippe Uzan, ``Bouncing compact objects. {II}.
  effective theory of a pulsating planck star,''
  \href{http://dx.doi.org/10.1103/physrevd.102.124041}{{\em Physical Review D}
  {\bfseries 102} no.~12, (Dec, 2020) }.
  \url{https://doi.org/10.1103%2Fphysrevd.102.124041}.

\bibitem{chandrasekhar1976solution}
Chandrasekhar, Subrahmanyan, ``The solution of dirac's equation in kerr
  geometry,'' {\em Proceedings of the Royal Society of London. A. Mathematical
  and Physical Sciences} {\bfseries 349} no.~1659, (1976) 571--575.

\bibitem{page1976dirac}
Page, Don N, ``Dirac equation around a charged, rotating black hole,'' {\em
  Physical Review D} {\bfseries 14} no.~6, (1976) 1509.

\bibitem{Teukolsky}
{Teukolsky}, Saul A., ``{Perturbations of a Rotating Black Hole. I. Fundamental
  Equations for Gravitational, Electromagnetic, and Neutrino-Field
  Perturbations},'' \href{http://dx.doi.org/10.1086/152444}{{\em \apj}
  {\bfseries 185} (Oct., 1973) 635--648}.

\bibitem{NP}
Newman,Ezra and Penrose,Roger, ``An approach to gravitational radiation by a
  method of spin coefficients,''
  \href{http://dx.doi.org/10.1063/1.1724257}{{\em Journal of Mathematical
  Physics} {\bfseries 3} no.~3, (1962) 566--578},
  \href{http://arxiv.org/abs/https://doi.org/10.1063/1.1724257}{{\ttfamily
  https://doi.org/10.1063/1.1724257}}. \url{https://doi.org/10.1063/1.1724257}.

\bibitem{Arbey:2021jif}
Arbey, Alexandre and Auffinger, J\'er\'emy and Geiller, Marc and Livine, Etera
  R. and Sartini, Francesco, ``{Hawking radiation by spherically-symmetric
  static black holes for all spins: Teukolsky equations and potentials},''
  \href{http://dx.doi.org/10.1103/PhysRevD.103.104010}{{\em Phys. Rev. D}
  {\bfseries 103} no.~10, (2021) 104010},
  \href{http://arxiv.org/abs/2101.02951}{{\ttfamily arXiv:2101.02951 [gr-qc]}}.

\bibitem{wkb1}
Iyer, Sai and Will, Clifford M., ``Black-hole normal modes: A wkb approach. i.
  foundations and application of a higher-order wkb analysis of
  potential-barrier scattering,''
  \href{http://dx.doi.org/10.1103/PhysRevD.35.3621}{{\em Phys. Rev. D}
  {\bfseries 35} (Jun, 1987) 3621--3631}.
  \url{https://link.aps.org/doi/10.1103/PhysRevD.35.3621}.

\bibitem{Konoplya_2019}
R A Konoplya and A Zhidenko and A F Zinhailo, ``Higher order {WKB} formula for
  quasinormal modes and grey-body factors: recipes for quick and accurate
  calculations,'' \href{http://dx.doi.org/10.1088/1361-6382/ab2e25}{{\em
  Classical and Quantum Gravity} {\bfseries 36} no.~15, (Jul, 2019) 155002}.
  \url{https://doi.org/10.1088%2F1361-6382%2Fab2e25}.

\bibitem{Cho:2003qe}
Cho, Hing Tong, ``{Dirac quasinormal modes in Schwarzschild black hole
  space-times},'' \href{http://dx.doi.org/10.1103/PhysRevD.68.024003}{{\em
  Phys. Rev. D} {\bfseries 68} (2003) 024003},
  \href{http://arxiv.org/abs/gr-qc/0303078}{{\ttfamily arXiv:gr-qc/0303078}}.

\bibitem{1973ApJ...185..649P}
{Press}, William H. and {Teukolsky}, Saul A., ``{Perturbations of a Rotating
  Black Hole. II. Dynamical Stability of the Kerr Metric},''
  \href{http://dx.doi.org/10.1086/152445}{{\em \apj} {\bfseries 185} (Oct.,
  1973) 649--674}.

\bibitem{1974ApJ...193..443T}
{Teukolsky}, S.~A. and {Press}, W.~H., ``{Perturbations of a rotating black
  hole. III. Interaction of the hole with gravitational and electromagnetic
  radiation.},'' \href{http://dx.doi.org/10.1086/153180}{{\em \apj} {\bfseries
  193} (Oct., 1974) 443--461}.

\bibitem{PhysRevD.13.198}
Page, Don N., ``Particle emission rates from a black hole: Massless particles
  from an uncharged, nonrotating hole,''
  \href{http://dx.doi.org/10.1103/PhysRevD.13.198}{{\em Phys. Rev. D}
  {\bfseries 13} (Jan, 1976) 198--206}.
  \url{https://link.aps.org/doi/10.1103/PhysRevD.13.198}.

\bibitem{PhysRevD.14.3260}
Page, Don N., ``Particle emission rates from a black hole. ii. massless
  particles from a rotating hole,''
  \href{http://dx.doi.org/10.1103/PhysRevD.14.3260}{{\em Phys. Rev. D}
  {\bfseries 14} (Dec, 1976) 3260--3273}.
  \url{https://link.aps.org/doi/10.1103/PhysRevD.14.3260}.

\bibitem{PhysRevD.16.2402}
Page, Don N., ``Particle emission rates from a black hole. iii. charged leptons
  from a nonrotating hole,''
  \href{http://dx.doi.org/10.1103/PhysRevD.16.2402}{{\em Phys. Rev. D}
  {\bfseries 16} (Oct, 1977) 2402--2411}.
  \url{https://link.aps.org/doi/10.1103/PhysRevD.16.2402}.

\bibitem{1975RSPSA.345..145C}
{Chandrasekhar}, S. and {Detweiler}, S., ``{On the Equations Governing the
  Axisymmetric Perturbations of the Kerr Black Hole},''
  \href{http://dx.doi.org/10.1098/rspa.1975.0130}{{\em Proceedings of the Royal
  Society of London Series A} {\bfseries 345} no.~1641, (Aug., 1975) 145--167}.

\bibitem{1976RSPSA.348...39C}
{Chandrasekhar}, S., ``{On a Transformation of Teukolsky's Equation and the
  Electromagnetic Perturbations of the Kerr Black Hole},''
  \href{http://dx.doi.org/10.1098/rspa.1976.0022}{{\em Proceedings of the Royal
  Society of London Series A} {\bfseries 348} no.~1652, (Feb., 1976) 39--55}.

\bibitem{1976RSPSA.350..165C}
{Chandrasekhar}, S. and {Detweiler}, S., ``{On the Equations Governing the
  Gravitational Perturbations of the Kerr Black Hole},''
  \href{http://dx.doi.org/10.1098/rspa.1976.0101}{{\em Proceedings of the Royal
  Society of London Series A} {\bfseries 350} no.~1661, (Aug., 1976) 165--174}.

\bibitem{1977RSPSA.352..325C}
{Chandrasekhar}, S. and {Detweiler}, S., ``{On the Reflexion and Transmission
  of Neutrino Waves by a Kerr Black Hole},''
  \href{http://dx.doi.org/10.1098/rspa.1977.0002}{{\em Proceedings of the Royal
  Society of London Series A} {\bfseries 352} no.~1670, (Jan., 1977) 325--338}.

\bibitem{universe8010050}
Alonso-Serrano, Ana and Liska, Marek, ``Quantum gravity phenomenology from the
  thermodynamics of spacetime,''
  \href{http://dx.doi.org/10.3390/universe8010050}{{\em Universe} {\bfseries 8}
  no.~1, (2022) }. \url{https://www.mdpi.com/2218-1997/8/1/50}.

\bibitem{Mele:2021hro}
Mele, Fabio M. and M\"unch, Johannes and Pateloudis, Stratos, ``{Quantum
  corrected polymer black hole thermodynamics: mass relations and logarithmic
  entropy correction},''
  \href{http://dx.doi.org/10.1088/1475-7516/2022/02/011}{{\em JCAP} {\bfseries
  02} no.~02, (2022) 011}, \href{http://arxiv.org/abs/2102.04788}{{\ttfamily
  arXiv:2102.04788 [gr-qc]}}.

\bibitem{PierreAuger:2021mve}
{\bfseries Pierre Auger} Collaboration, Abreu, Pedro and others,
  ``{Constraining Lorentz Invariance Violation using the muon content of
  extensive air showers measured at the Pierre Auger Observatory},''
  \href{http://dx.doi.org/10.22323/1.395.0340}{{\em PoS} {\bfseries ICRC2021}
  (2021) 340}.

\bibitem{Lobo:2021yem}
Lobo, Iarley P. and Pfeifer, Christian and Morais, Pedro H. and Batista, Rafael
  Alves and Bezerra, Valdir B., ``{Two-body decays in deformed relativity},''
  \href{http://dx.doi.org/10.1007/JHEP09(2022)003}{{\em JHEP} {\bfseries 09}
  (2022) 003}, \href{http://arxiv.org/abs/2112.12172}{{\ttfamily
  arXiv:2112.12172 [hep-ph]}}.

\bibitem{Lobo:2020qoa}
Lobo, Iarley P. and Pfeifer, Christian, ``{Reaching the Planck scale with muon
  lifetime measurements},''
  \href{http://dx.doi.org/10.1103/PhysRevD.103.106025}{{\em Phys. Rev. D}
  {\bfseries 103} no.~10, (2021) 106025},
  \href{http://arxiv.org/abs/2011.10069}{{\ttfamily arXiv:2011.10069
  [hep-ph]}}.

\end{thebibliography}\endgroup

\appendix

\section{Tables of quasinormal frequencies}
In the following, we list the quasinormal frequencies calculated using the 6th order WKB method and used throughout this paper for different values of the (de-)amplification parameter $\beta$, mode number $n$, multipole number $l$ and LQG parameters $\lambda_1$ and $\lambda_2$ (we have fixed $M_{BH}=1$).
\subsection{$\beta=\frac{3}{5}$}

\subsubsection{$l=\frac{1}{2}$}

\begin{equation}    
    \begin{tabular}{|c|c|c|c|c|}
\hline
 $l=1/2$ &  &  &  &  \\
 \hline
 \text{$\lambda_{1}$} & \text{$\lambda_{2}$} & n=0 & n=1 & n=2 \\ \hline
 
 \text{$\lambda_{1} $=0} & \text{$\lambda_{2} $=0} & 0.182646\, -0.0949348 i & 0.144749\, -0.314329 i & 0.125138\, -0.568691 i \\ \hline
 \text{$\lambda_{1} $=0.05} & \text{$\lambda_{2} $=0.05} & 0.184713\, -0.0953346 i & 0.147266\, -0.315089 i & 0.127799\, -0.569211 i \\ 
  & \text{$\lambda_{2} $=0.10} & 0.185595\, -0.0955002 i & 0.148344\, -0.315398 i & 0.128941\, -0.569407 i \\
  & \text{$\lambda_{2} $=0.15} & 0.186282\, -0.0956272 i & 0.149185\, -0.315632 i & 0.129833\, -0.569549 i \\
  & \text{$\lambda_{2} $=0.20} & 0.186869\, -0.0957341 i & 0.149904\, -0.315827 i & 0.130596\, -0.569662 i \\
  & \text{$\lambda_{2} $=0.25} & 0.187391\, -0.0958282 i & 0.150545\, -0.315997 i & 0.131277\, -0.569757 i \\
  & \text{$\lambda_{2} $=0.30} & 0.187868\, -0.0959132 i & 0.151131\, -0.31615 i & 0.1319\, -0.569839 i \\ \hline
 \text{$\lambda_{1} $=0.10} & \text{$\lambda_{2} $=0.05} & 0.185595\, -0.0955002 i & 0.148344\, -0.315398 i & 0.128941\, -0.569407 i \\
  & \text{$\lambda_{2} $=0.10} & 0.186869\, -0.0957341 i & 0.149904\, -0.315827 i & 0.130596\, -0.569662 i \\
  & \text{$\lambda_{2} $=0.15} & 0.187868\, -0.0959132 i & 0.151131\, -0.31615 i & 0.1319\, -0.569839 i \\
  & \text{$\lambda_{2} $=0.20} & 0.188727\, -0.0960639 i & 0.152186\, -0.316417 i & 0.133022\, -0.569975 i \\
  & \text{$\lambda_{2} $=0.25} & 0.189495\, -0.0961963 i & 0.153132\, -0.316649 i & 0.13403\, -0.570084 i \\
  & \text{$\lambda_{2} $=0.30} & 0.1902\, -0.0963157 i & 0.154001\, -0.316855 i & 0.134956\, -0.570174 i \\ \hline
 \text{$\lambda_{1} $=0.15} & \text{$\lambda_{2} $=0.05} & 0.186282\, -0.0956272 i & 0.149185\, -0.315632 i & 0.129833\, -0.569549 i \\ 
  & \text{$\lambda_{2} $=0.10} & 0.187868\, -0.0959132 i & 0.151131\, -0.31615 i & 0.1319\, -0.569839 i \\
  & \text{$\lambda_{2} $=0.15} & 0.18912\, -0.096132 i & 0.152671\, -0.316537 i & 0.133538\, -0.570033 i \\
  & \text{$\lambda_{2} $=0.20} & 0.1902\, -0.0963157 i & 0.154001\, -0.316855 i & 0.134956\, -0.570174 i \\
  & \text{$\lambda_{2} $=0.25} & 0.191171\, -0.0964768 i & 0.1552\, -0.317129 i & 0.136233\, -0.570281 i \\
  & \text{$\lambda_{2} $=0.30} & 0.192065\, -0.0966218 i & 0.156305\, -0.317371 i & 0.137413\, -0.570362 i \\ \hline
 \text{$\lambda_{1} $=0.20} & \text{$\lambda_{2} $=0.05} & 0.186869\, -0.0957341 i & 0.149904\, -0.315827 i & 0.130596\, -0.569662 i \\
  & \text{$\lambda_{2} $=0.10} & 0.188727\, -0.0960639 i & 0.152186\, -0.316417 i & 0.133022\, -0.569975 i \\
  & \text{$\lambda_{2} $=0.15} & 0.1902\, -0.0963157 i & 0.154001\, -0.316855 i & 0.134956\, -0.570174 i \\
  & \text{$\lambda_{2} $=0.20} & 0.191476\, -0.0965267 i & 0.155577\, -0.317213 i & 0.136636\, -0.57031 i \\
  & \text{$\lambda_{2} $=0.25} & 0.192627\, -0.0967114 i & 0.157001\, -0.317518 i & 0.138156\, -0.570405 i \\
  & \text{$\lambda_{2} $=0.30} & 0.19369\, -0.0968773 i & 0.158319\, -0.317786 i & 0.139564\, -0.570469 i \\ \hline
 \text{$\lambda_{1} $=0.25} & \text{$\lambda_{2} $=0.05} & 0.187391\, -0.0958282 i & 0.150545\, -0.315997 i & 0.131277\, -0.569757 i \\
  & \text{$\lambda_{2} $=0.10} & 0.189495\, -0.0961963 i & 0.153132\, -0.316649 i & 0.13403\, -0.570084 i \\
  & \text{$\lambda_{2} $=0.15} & 0.191171\, -0.0964768 i & 0.1552\, -0.317129 i & 0.136233\, -0.570281 i \\
  & \text{$\lambda_{2} $=0.20} & 0.192627\, -0.0967114 i & 0.157001\, -0.317518 i & 0.138156\, -0.570405 i \\
  & \text{$\lambda_{2} $=0.25} & 0.193945\, -0.0969164 i & 0.158635\, -0.317848 i & 0.139902\, -0.57048 i \\
  & \text{$\lambda_{2} $=0.30} & 0.195166\, -0.0971 i & 0.160152\, -0.318134 i & 0.141523\, -0.57052 i \\ \hline
 \text{$\lambda_{1} $=0.30} & \text{$\lambda_{2} $=0.05} & 0.187868\, -0.0959132 i & 0.151131\, -0.31615 i & 0.1319\, -0.569839 i \\
  & \text{$\lambda_{2} $=0.10} & 0.1902\, -0.0963157 i & 0.154001\, -0.316855 i & 0.134956\, -0.570174 i \\
  & \text{$\lambda_{2} $=0.15} & 0.192065\, -0.0966218 i & 0.156305\, -0.317371 i & 0.137413\, -0.570362 i \\
  & \text{$\lambda_{2} $=0.20} & 0.19369\, -0.0968773 i & 0.158319\, -0.317786 i & 0.139564\, -0.570469 i \\
  & \text{$\lambda_{2} $=0.25} & 0.195166\, -0.0971 i & 0.160152\, -0.318134 i & 0.141523\, -0.57052 i \\
  & \text{$\lambda_{2} $=0.30} & 0.196537\, -0.097299 i & 0.161858\, -0.318434 i & 0.143348\, -0.570528 i \\ \hline

\end{tabular}
\end{equation}

\subsubsection{$l=\frac{3}{2}$}

\begin{equation}    
    \begin{tabular}{|c|c|c|c|c|}
\hline
 $l=3/2$ &  &  &  &  \\
 \hline
 \text{$\lambda_{1}$} & \text{$\lambda_{2}$} & n=0 & n=1 & n=2 \\ \hline
 
 \text{$\lambda_{1} $=0} & \text{$\lambda_{2} $=0} & 0.380068\, -0.0963659 i & 0.355857\, -0.297271 i & 0.318931\, -0.518572 i \\ \hline
 \text{$\lambda_{1} $=0.05} & \text{$\lambda_{2} $=0.05} & 0.384166\, -0.0966904 i & 0.360245\, -0.298136 i & 0.323753\, -0.519676 i \\ 
  & \text{$\lambda_{2} $=0.10} & 0.385912\, -0.0968238 i & 0.362117\, -0.29849 i & 0.325811\, -0.520121 i \\
  & \text{$\lambda_{2} $=0.15} & 0.387273\, -0.0969256 i & 0.363576\, -0.298759 i & 0.327416\, -0.520457 i \\
  & \text{$\lambda_{2} $=0.20} & 0.388435\, -0.0970111 i & 0.364822\, -0.298984 i & 0.328787\, -0.520736 i \\
  & \text{$\lambda_{2} $=0.25} & 0.389469\, -0.097086 i & 0.365931\, -0.299181 i & 0.330008\, -0.520979 i \\
  & \text{$\lambda_{2} $=0.30} & 0.390413\, -0.0971535 i & 0.366944\, -0.299358 i & 0.331124\, -0.521196 i \\ \hline
  \text{$\lambda_{1} $=0.10} & \text{$\lambda_{2} $=0.05} & 0.385912\, -0.0968238 i & 0.362117\, -0.29849 i & 0.325811\, -0.520121 i \\
  & \text{$\lambda_{2} $=0.10} & 0.388435\, -0.0970111 i & 0.364822\, -0.298984 i & 0.328787\, -0.520736 i \\
  & \text{$\lambda_{2} $=0.15} & 0.390413\, -0.0971535 i & 0.366944\, -0.299358 i & 0.331124\, -0.521196 i \\
  & \text{$\lambda_{2} $=0.20} & 0.392112\, -0.0972727 i & 0.368766\, -0.29967 i & 0.33313\, -0.521574 i \\
  & \text{$\lambda_{2} $=0.25} & 0.393631\, -0.0973768 i & 0.370397\, -0.299941 i & 0.334928\, -0.5219 i \\
  & \text{$\lambda_{2} $=0.30} & 0.395025\, -0.0974702 i & 0.371894\, -0.300183 i & 0.336576\, -0.522187 i \\ \hline
  \text{$\lambda_{1} $=0.15} & \text{$\lambda_{2} $=0.05} & 0.387273\, -0.0969256 i & 0.363576\, -0.298759 i & 0.327416\, -0.520457 i \\
  & \text{$\lambda_{2} $=0.10} & 0.390413\, -0.0971535 i & 0.366944\, -0.299358 i & 0.331124\, -0.521196 i \\
  & \text{$\lambda_{2} $=0.15} & 0.39289\, -0.0973263 i & 0.369601\, -0.29981 i & 0.334051\, -0.521742 i \\
  & \text{$\lambda_{2} $=0.20} & 0.395025\, -0.0974702 i & 0.371894\, -0.300183 i & 0.336576\, -0.522187 i \\
  & \text{$\lambda_{2} $=0.25} & 0.396943\, -0.0975956 i & 0.373954\, -0.300507 i & 0.338847\, -0.522566 i \\
  & \text{$\lambda_{2} $=0.30} & 0.398709\, -0.0977075 i & 0.37585\, -0.300795 i & 0.340938\, -0.522897 i \\
  \hline
  \text{$\lambda_{1} $=0.20} & \text{$\lambda_{2} $=0.05} & 0.388435\, -0.0970111 i & 0.364822\, -0.298984 i & 0.328787\, -0.520736 i \\
  & \text{$\lambda_{2} $=0.10} & 0.392112\, -0.0972727 i & 0.368766\, -0.29967 i & 0.33313\, -0.521574 i \\
  & \text{$\lambda_{2} $=0.15} & 0.395025\, -0.0974702 i & 0.371894\, -0.300183 i & 0.336576\, -0.522187 i \\
  & \text{$\lambda_{2} $=0.20} & 0.397547\, -0.0976342 i & 0.374602\, -0.300607 i & 0.339561\, -0.522681 i \\
  & \text{$\lambda_{2} $=0.25} & 0.39982\, -0.0977764 i & 0.377045\, -0.300971 i & 0.342255\, -0.523096 i \\
  & \text{$\lambda_{2} $=0.30} & 0.40192\, -0.0979028 i & 0.379301\, -0.301292 i & 0.344743\, -0.523454 i \\
  \hline
  \text{$\lambda_{1} $=0.25} & \text{$\lambda_{2} $=0.05} & 0.389469\, -0.097086 i & 0.365931\, -0.299181 i & 0.330008\, -0.520979 i \\
  & \text{$\lambda_{2} $=0.10} & 0.393631\, -0.0973768 i & 0.370397\, -0.299941 i & 0.334928\, -0.5219 i \\
  & \text{$\lambda_{2} $=0.15} & 0.396943\, -0.0975956 i & 0.373954\, -0.300507 i & 0.338847\, -0.522566 i \\
  & \text{$\lambda_{2} $=0.20} & 0.39982\, -0.0977764 i & 0.377045\, -0.300971 i & 0.342255\, -0.523096 i \\
  & \text{$\lambda_{2} $=0.25} & 0.402423\, -0.0979324 i & 0.379842\, -0.301367 i & 0.34534\, -0.523536 i \\
  & \text{$\lambda_{2} $=0.30} & 0.404833\, -0.0980705 i & 0.382433\, -0.301714 i & 0.348198\, -0.52391 i \\
  \hline
 \text{$\lambda_{1} $=0.30} & \text{$\lambda_{2} $=0.05} & 0.390413\, -0.0971535 i & 0.366944\, -0.299358 i & 0.331124\, -0.521196 i \\
  & \text{$\lambda_{2} $=0.10} & 0.395025\, -0.0974702 i & 0.371894\, -0.300183 i & 0.336576\, -0.522187 i \\
  & \text{$\lambda_{2} $=0.15} & 0.398709\, -0.0977075 i & 0.37585\, -0.300795 i & 0.340938\, -0.522897 i \\
  & \text{$\lambda_{2} $=0.20} & 0.40192\, -0.0979028 i & 0.379301\, -0.301292 i & 0.344743\, -0.523454 i \\
  & \text{$\lambda_{2} $=0.25} & 0.404833\, -0.0980705 i & 0.382433\, -0.301714 i & 0.348198\, -0.52391 i \\
  & \text{$\lambda_{2} $=0.30} & 0.407538\, -0.098218 i & 0.385343\, -0.30208 i & 0.351409\, -0.52429 i \\ \hline

\end{tabular}
\end{equation}

\subsubsection{$l=\frac{5}{2}$}

\begin{equation}    
    \begin{tabular}{|c|c|c|c|c|}
\hline
 $l=5/2$ &  &  &  &  \\
 \hline
 \text{$\lambda_{1}$} & \text{$\lambda_{2}$} & n=0 & n=1 & n=2 \\ \hline
 
 \text{$\lambda_{1}$=0} & \text{$\lambda_{2} $=0} & 0.574094\, -0.096307 i & 0.557016\, -0.292717 i & 0.526534\, -0.499713 i \\ \hline
 \text{$\lambda_{1} $=0.05} & \text{$\lambda_{2} $=0.05} & 0.58021\, -0.0966355 i & 0.56334\, -0.293653 i & 0.53323\, -0.501106 i \\
  & \text{$\lambda_{2} $=0.10} & 0.582818\, -0.0967704 i & 0.566038\, -0.294036 i & 0.536087\, -0.501673 i \\
  & \text{$\lambda_{2} $=0.15} & 0.58485\, -0.0968734 i & 0.568139\, -0.294328 i & 0.538314\, -0.502103 i \\
  & \text{$\lambda_{2} $=0.20} & 0.586584\, -0.0969597 i & 0.569933\, -0.294572 i & 0.540215\, -0.502463 i \\
  & \text{$\lambda_{2} $=0.25} & 0.588128\, -0.0970355 i & 0.571531\, -0.294786 i & 0.541908\, -0.502777 i \\
  & \text{$\lambda_{2} $=0.30} & 0.589538\, -0.0971037 i & 0.57299\, -0.294979 i & 0.543455\, -0.503058 i \\ \hline
 \text{$\lambda_{1} $=0.10} & \text{$\lambda_{2} $=0.05} & 0.582818\, -0.0967704 i & 0.566038\, -0.294036 i & 0.536087\, -0.501673 i \\
  & \text{$\lambda_{2} $=0.10} & 0.586584\, -0.0969597 i & 0.569933\, -0.294572 i & 0.540215\, -0.502463 i \\
  & \text{$\lambda_{2} $=0.15} & 0.589538\, -0.0971037 i & 0.57299\, -0.294979 i & 0.543455\, -0.503058 i \\
  & \text{$\lambda_{2} $=0.20} & 0.592073\, -0.097224 i & 0.575614\, -0.295318 i & 0.546237\, -0.503553 i \\
  & \text{$\lambda_{2} $=0.25} & 0.594342\, -0.0973291 i & 0.577962\, -0.295614 i & 0.548727\, -0.503982 i \\
  & \text{$\lambda_{2} $=0.30} & 0.596422\, -0.0974234 i & 0.580116\, -0.295879 i & 0.551011\, -0.504364 i \\
  \hline
 \text{$\lambda_{1} $=0.15} & \text{$\lambda_{2} $=0.05} & 0.58485\, -0.0968734 i & 0.568139\, -0.294328 i & 0.538314\, -0.502103 i \\
  & \text{$\lambda_{2} $=0.10} & 0.589538\, -0.0971037 i & 0.57299\, -0.294979 i & 0.543455\, -0.503058 i \\
  & \text{$\lambda_{2} $=0.15} & 0.593235\, -0.0972781 i & 0.576816\, -0.295471 i & 0.547512\, -0.503774 i \\
  & \text{$\lambda_{2} $=0.20} & 0.596422\, -0.0974234 i & 0.580116\, -0.295879 i & 0.551011\, -0.504364 i \\
  & \text{$\lambda_{2} $=0.25} & 0.599286\, -0.0975498 i & 0.583081\, -0.296233 i & 0.554157\, -0.504873 i \\
  & \text{$\lambda_{2} $=0.30} & 0.601922\, -0.0976627 i & 0.585811\, -0.296548 i & 0.557053\, -0.505323 i \\
  \hline
 \text{$\lambda_{1} $=0.20} & \text{$\lambda_{2} $=0.05} & 0.586584\, -0.0969597 i & 0.569933\, -0.294572 i & 0.540215\, -0.502463 i \\
  & \text{$\lambda_{2} $=0.10} & 0.592073\, -0.097224 i & 0.575614\, -0.295318 i & 0.546237\, -0.503553 i \\
  & \text{$\lambda_{2} $=0.15} & 0.596422\, -0.0974234 i & 0.580116\, -0.295879 i & 0.551011\, -0.504364 i \\
  & \text{$\lambda_{2} $=0.20} & 0.600187\, -0.0975888 i & 0.584014\, -0.296342 i & 0.555146\, -0.505028 i \\
  & \text{$\lambda_{2} $=0.25} & 0.603581\, -0.0977321 i & 0.587529\, -0.296742 i & 0.558876\, -0.505597 i \\
  & \text{$\lambda_{2} $=0.30} & 0.606715\, -0.0978595 i & 0.590775\, -0.297096 i & 0.562321\, -0.506097 i \\
  \hline
 \text{$\lambda_{1} $=0.25} & \text{$\lambda_{2} $=0.05} & 0.588128\, -0.0970355 i & 0.571531\, -0.294786 i & 0.541908\, -0.502777 i \\
  & \text{$\lambda_{2} $=0.10} & 0.594342\, -0.0973291 i & 0.577962\, -0.295614 i & 0.548727\, -0.503982 i \\
  & \text{$\lambda_{2} $=0.15} & 0.599286\, -0.0975498 i & 0.583081\, -0.296233 i & 0.554157\, -0.504873 i \\
  & \text{$\lambda_{2} $=0.20} & 0.603581\, -0.0977321 i & 0.587529\, -0.296742 i & 0.558876\, -0.505597 i \\
  & \text{$\lambda_{2} $=0.25} & 0.607466\, -0.0978893 i & 0.591553\, -0.297179 i & 0.563147\, -0.506213 i \\
  & \text{$\lambda_{2} $=0.30} & 0.611064\, -0.0980284 i & 0.595281\, -0.297564 i & 0.567104\, -0.506748 i \\
  \hline
 \text{$\lambda_{1} $=0.30} & \text{$\lambda_{2} $=0.05} & 0.589538\, -0.0971037 i & 0.57299\, -0.294979 i & 0.543455\, -0.503058 i \\
  & \text{$\lambda_{2} $=0.10} & 0.596422\, -0.0974234 i & 0.580116\, -0.295879 i & 0.551011\, -0.504364 i \\
  & \text{$\lambda_{2} $=0.15} & 0.601922\, -0.0976627 i & 0.585811\, -0.296548 i & 0.557053\, -0.505323 i \\
  & \text{$\lambda_{2} $=0.20} & 0.606715\, -0.0978595 i & 0.590775\, -0.297096 i & 0.562321\, -0.506097 i \\
  & \text{$\lambda_{2} $=0.25} & 0.611064\, -0.0980284 i & 0.595281\, -0.297564 i & 0.567104\, -0.506748 i \\
  & \text{$\lambda_{2} $=0.30} & 0.615102\, -0.098177 i & 0.599466\, -0.297972 i & 0.571547\, -0.50731 i \\
  \hline

\end{tabular}
\end{equation}

\subsection{$\beta=\frac{5}{3}$}

\subsubsection{$l=\frac{1}{2}$}

\begin{equation}    
    \begin{tabular}{|c|c|c|c|c|}
\hline
 $l=1/2$ &  &  &  &  \\
 \hline
 \text{$\lambda_{1}$} & \text{$\lambda_{2}$} & n=0 & n=1 & n=2 \\ \hline
  \text{$\lambda_{1} $=0} & \text{$\lambda_{2} $=0} & 0.182646\, -0.0949348 i & 0.144749\, -0.314329 i & 0.125138\, -0.568691 i \\ \hline
 \text{$\lambda_{1} $=0.05} & \text{$\lambda_{2} $=0.05} & 0.183628\, -0.0951274 i & 0.145944\, -0.314699 i & 0.126401\, -0.568954 i \\
  & \text{$\lambda_{2} $=0.10} & 0.18404\, -0.0952079 i & 0.146447\, -0.314855 i & 0.126934\, -0.569066 i \\
  & \text{$\lambda_{2} $=0.15} & 0.184358\, -0.0952703 i & 0.146836\, -0.314976 i & 0.127348\, -0.569156 i \\
  & \text{$\lambda_{2} $=0.20} & 0.184627\, -0.0953233 i & 0.147165\, -0.31508 i & 0.1277\, -0.569235 i \\
  & \text{$\lambda_{2} $=0.25} & 0.184865\, -0.0953705 i & 0.147458\, -0.315174 i & 0.128012\, -0.569308 i \\
  & \text{$\lambda_{2} $=0.30} & 0.185081\, -0.0954134 i & 0.147723\, -0.31526 i & 0.128297\, -0.569376 i \\
  \hline
 \text{$\lambda_{1} $=0.10} & \text{$\lambda_{2} $=0.05} & 0.18404\, -0.0952079 i & 0.146447\, -0.314855 i & 0.126934\, -0.569066 i \\
  & \text{$\lambda_{2} $=0.10} & 0.184627\, -0.0953233 i & 0.147165\, -0.31508 i & 0.1277\, -0.569235 i \\
  & \text{$\lambda_{2} $=0.15} & 0.185081\, -0.0954134 i & 0.147723\, -0.31526 i & 0.128297\, -0.569377 i \\
  & \text{$\lambda_{2} $=0.20} & 0.185466\, -0.0954906 i & 0.148197\, -0.315416 i & 0.128807\, -0.569506 i \\
  & \text{$\lambda_{2} $=0.25} & 0.185807\, -0.0955598 i & 0.148619\, -0.315559 i & 0.129262\, -0.56963 i \\
  & \text{$\lambda_{2} $=0.30} & 0.186117\, -0.0956234 i & 0.149002\, -0.315693 i & 0.129678\, -0.569751 i \\
  \hline
 \text{$\lambda_{1} $=0.15} & \text{$\lambda_{2} $=0.05} & 0.184358\, -0.0952703 i & 0.146836\, -0.314976 i & 0.127348\, -0.569156 i \\
  & \text{$\lambda_{2} $=0.10} & 0.185081\, -0.0954134 i & 0.147723\, -0.31526 i & 0.128297\, -0.569377 i \\
  & \text{$\lambda_{2} $=0.15} & 0.185641\, -0.0955261 i & 0.148414\, -0.315489 i & 0.12904\, -0.569568 i \\
  & \text{$\lambda_{2} $=0.20} & 0.186117\, -0.0956234 i & 0.149002\, -0.315693 i & 0.129678\, -0.569751 i \\
  & \text{$\lambda_{2} $=0.25} & 0.186538\, -0.0957113 i & 0.149525\, -0.315881 i & 0.130249\, -0.56993 i \\
  & \text{$\lambda_{2} $=0.30} & 0.18692\, -0.0957926 i & 0.150002\, -0.31606 i & 0.130774\, -0.570109 i \\
  \hline
 \text{$\lambda_{1} $=0.20} & \text{$\lambda_{2} $=0.05} & 0.184627\, -0.0953233 i & 0.147165\, -0.31508 i & 0.1277\, -0.569235 i \\
  & \text{$\lambda_{2} $=0.10} & 0.185466\, -0.0954906 i & 0.148197\, -0.315416 i & 0.128807\, -0.569506 i \\
  & \text{$\lambda_{2} $=0.15} & 0.186117\, -0.0956234 i & 0.149002\, -0.315693 i & 0.129678\, -0.569751 i \\
  & \text{$\lambda_{2} $=0.20} & 0.186669\, -0.095739 i & 0.149689\, -0.315942 i & 0.130429\, -0.569989 i \\
  & \text{$\lambda_{2} $=0.25} & 0.187158\, -0.0958441 i & 0.1503\, -0.316176 i & 0.131103\, -0.570229 i \\
  & \text{$\lambda_{2} $=0.30} & 0.187603\, -0.0959422 i & 0.150859\, -0.316401 i & 0.131724\, -0.570473 i \\
  \hline
 \text{$\lambda_{1} $=0.25} & \text{$\lambda_{2} $=0.05} & 0.184865\, -0.0953705 i & 0.147458\, -0.315174 i & 0.128012\, -0.569308 i \\
  & \text{$\lambda_{2} $=0.10} & 0.185807\, -0.0955598 i & 0.148619\, -0.315559 i & 0.129262\, -0.56963 i \\
  & \text{$\lambda_{2} $=0.15} & 0.186538\, -0.0957113 i & 0.149525\, -0.315881 i & 0.130249\, -0.56993 i \\
  & \text{$\lambda_{2} $=0.20} & 0.187158\, -0.0958441 i & 0.1503\, -0.316176 i & 0.131103\, -0.570229 i \\
  & \text{$\lambda_{2} $=0.25} & 0.187708\, -0.0959658 i & 0.150992\, -0.316456 i & 0.131873\, -0.570535 i \\
  & \text{$\lambda_{2} $=0.30} & 0.188207\, -0.0960801 i & 0.151623\, -0.316729 i & 0.132584\, -0.57085 i \\
  \hline
 \text{$\lambda_{1} $=0.30} & \text{$\lambda_{2} $=0.05} & 0.185081\, -0.0954134 i & 0.147723\, -0.31526 i & 0.128297\, -0.569376 i \\
  & \text{$\lambda_{2} $=0.10} & 0.186117\, -0.0956234 i & 0.149002\, -0.315693 i & 0.129678\, -0.569751 i \\
  & \text{$\lambda_{2} $=0.15} & 0.18692\, -0.0957926 i & 0.150002\, -0.31606 i & 0.130774\, -0.570109 i \\
  & \text{$\lambda_{2} $=0.20} & 0.187603\, -0.0959422 i & 0.150859\, -0.316401 i & 0.131724\, -0.570473 i \\
  & \text{$\lambda_{2} $=0.25} & 0.188207\, -0.0960801 i & 0.151623\, -0.316729 i & 0.132584\, -0.57085 i \\
  & \text{$\lambda_{2} $=0.30} & 0.188756\, -0.0962104 i & 0.152322\, -0.317051 i & 0.133379\, -0.571243 i \\
  \hline

\end{tabular}
\end{equation}

\subsubsection{$l=\frac{3}{2}$}

\begin{equation}    
    \begin{tabular}{|c|c|c|c|c|}
\hline
 $l=3/2$ &  &  &  &  \\
 \hline
 \text{$\lambda_{1}$} & \text{$\lambda_{2}$} & n=0 & n=1 & n=2 \\ \hline
 \text{$\lambda_{1} $=0} & \text{$\lambda_{2} $=0} & 0.380068\, -0.0963659 i & 0.355857\, -0.297271 i & 0.318931\, -0.518572 i \\ \hline
 \text{$\lambda_{1} $=0.05} & \text{$\lambda_{2} $=0.05} & 0.382016\, -0.0965222 i & 0.357943\, -0.297689 i & 0.321222\, -0.519107 i \\
  & \text{$\lambda $2=0.10} & 0.382834\, -0.0965868 i & 0.358819\, -0.297861 i & 0.322184\, -0.519327 i \\
  & \text{$\lambda $2=0.15} & 0.383466\, -0.0966363 i & 0.359496\, -0.297992 i & 0.322928\, -0.519494 i \\
  & \text{$\lambda $2=0.20} & 0.384002\, -0.0966779 i & 0.360069\, -0.298103 i & 0.323559\, -0.519635 i \\
  & \text{$\lambda $2=0.25} & 0.384476\, -0.0967146 i & 0.360578\, -0.2982 i & 0.324117\, -0.519758 i \\
  & \text{$\lambda $2=0.30} & 0.384907\, -0.0967477 i & 0.361039\, -0.298288 i & 0.324624\, -0.51987 i \\ \hline
 \text{$\lambda_{1} $=0.10} & \text{$\lambda_{2} $=0.05} & 0.382834\, -0.0965868 i & 0.358819\, -0.297861 i & 0.322184\, -0.519327 i \\
  & \text{$\lambda $2=0.10} & 0.384002\, -0.0966779 i & 0.360069\, -0.298103 i & 0.323559\, -0.519635 i \\
  & \text{$\lambda $2=0.15} & 0.384907\, -0.0967477 i & 0.361039\, -0.298288 i & 0.324624\, -0.51987 i \\
  & \text{$\lambda $2=0.20} & 0.385676\, -0.0968064 i & 0.361863\, -0.298444 i & 0.32553\, -0.520066 i \\
  & \text{$\lambda $2=0.25} & 0.386359\, -0.096858 i & 0.362594\, -0.298581 i & 0.326333\, -0.520238 i \\
  & \text{$\lambda $2=0.30} & 0.38698\, -0.0969047 i & 0.363259\, -0.298705 i & 0.327064\, -0.520393 i \\ \hline
 \text{$\lambda_{1} $=0.15} & \text{$\lambda_{2} $=0.05} & 0.383466\, -0.0966363 i & 0.359496\, -0.297992 i & 0.322928\, -0.519494 i \\
  & \text{$\lambda_{2} $=0.10} & 0.384907\, -0.0967477 i & 0.361039\, -0.298288 i & 0.324624\, -0.51987 i \\
  & \text{$\lambda_{2} $=0.15} & 0.386027\, -0.0968329 i & 0.362238\, -0.298515 i & 0.325942\, -0.520155 i \\
  & \text{$\lambda_{2} $=0.20} & 0.38698\, -0.0969047 i & 0.363259\, -0.298705 i & 0.327064\, -0.520393 i \\
  & \text{$\lambda_{2} $=0.25} & 0.387827\, -0.0969677 i & 0.364166\, -0.298871 i & 0.32806\, -0.520602 i \\
  & \text{$\lambda_{2} $=0.30} & 0.388599\, -0.0970246 i & 0.364993\, -0.299022 i & 0.328968\, -0.52079 i \\
  \hline
 \text{$\lambda_{1} $=0.20} & \text{$\lambda $2=0.05} & 0.384002\, -0.0966779 i & 0.360069\, -0.298103 i & 0.323559\, -0.519635 i \\
  & \text{$\lambda_{2} $=0.10} & 0.385676\, -0.0968064 i & 0.361863\, -0.298444 i & 0.32553\, -0.520066 i \\
  & \text{$\lambda_{2} $=0.15} & 0.38698\, -0.0969047 i & 0.363259\, -0.298705 i & 0.327064\, -0.520393 i \\
  & \text{$\lambda_{2} $=0.20} & 0.388092\, -0.0969873 i & 0.36445\, -0.298923 i & 0.328371\, -0.520667 i \\
  & \text{$\lambda_{2} $=0.25} & 0.389081\, -0.0970599 i & 0.365509\, -0.299115 i & 0.329535\, -0.520906 i \\
  & \text{$\lambda_{2} $=0.30} & 0.389983\, -0.0971254 i & 0.366475\, -0.299288 i & 0.330595\, -0.521121 i \\
  \hline
 \text{$\lambda_{1} $=0.25} & \text{$\lambda_{2} $=0.05} & 0.384476\, -0.0967146 i & 0.360578\, -0.2982 i & 0.324117\, -0.519758 i \\
  & \text{$\lambda_{2} $=0.10} & 0.386359\, -0.096858 i & 0.362594\, -0.298581 i & 0.326333\, -0.520238 i \\
  & \text{$\lambda_{2} $=0.15} & 0.387827\, -0.0969677 i & 0.364166\, -0.298871 i & 0.32806\, -0.520602 i \\
  & \text{$\lambda_{2} $=0.20} & 0.389081\, -0.0970599 i & 0.365509\, -0.299115 i & 0.329535\, -0.520906 i \\
  & \text{$\lambda_{2} $=0.25} & 0.390198\, -0.0971409 i & 0.366705\, -0.299329 i & 0.330848\, -0.521172 i \\
  & \text{$\lambda_{2} $=0.30} & 0.391218\, -0.0972139 i & 0.367796\, -0.299522 i & 0.332045\, -0.52141 i \\
  \hline
 \text{$\lambda_{1} $=0.30} & \text{$\lambda_{2} $=0.05} & 0.384907\, -0.0967477 i & 0.361039\, -0.298288 i & 0.324624\, -0.51987 i \\
  & \text{$\lambda_{2} $=0.10} & 0.38698\, -0.0969047 i & 0.363259\, -0.298705 i & 0.327064\, -0.520393 i \\
  & \text{$\lambda_{2} $=0.15} & 0.388599\, -0.0970246 i & 0.364993\, -0.299022 i & 0.328968\, -0.52079 i \\
  & \text{$\lambda_{2} $=0.20} & 0.389983\, -0.0971254 i & 0.366475\, -0.299288 i & 0.330595\, -0.521121 i \\
  & \text{$\lambda_{2} $=0.25} & 0.391218\, -0.0972139 i & 0.367796\, -0.299522 i & 0.332045\, -0.52141 i \\
  & \text{$\lambda_{2} $=0.30} & 0.392347\, -0.0972938 i & 0.369004\, -0.299732 i & 0.33337\, -0.52167 i \\
  \hline

\end{tabular}
\end{equation}

\subsubsection{$l=\frac{5}{2}$}

\begin{equation}    
    \begin{tabular}{|c|c|c|c|c|}
\hline
 $l=5/2$ &  &  &  &  \\
 \hline
 \text{$\lambda_{1}$} & \text{$\lambda_{2}$} & n=0 & n=1 & n=2 \\ \hline
 \text{$\lambda_{1} $=0} & \text{$\lambda_{2} $=0} & 0.574094\, -0.096307 i & 0.557016\, -0.292717 i & 0.526534\, -0.499713 i \\ \hline
 \text{$\lambda_{1} $=0.05} & \text{$\lambda_{2} $=0.05} & 0.577002\, -0.0964653 i & 0.560022\, -0.293169 i & 0.529716\, -0.500386 i \\
  & \text{$\lambda_{2} $=0.10} & 0.578223\, -0.0965306 i & 0.561285\, -0.293355 i & 0.531053\, -0.500663 i \\
  & \text{$\lambda_{2} $=0.15} & 0.579166\, -0.0965807 i & 0.56226\, -0.293497 i & 0.532086\, -0.500874 i \\
  & \text{$\lambda_{2} $=0.20} & 0.579966\, -0.0966228 i & 0.563088\, -0.293617 i & 0.532961\, -0.501052 i \\
  & \text{$\lambda_{2} $=0.25} & 0.580675\, -0.0966599 i & 0.56382\, -0.293722 i & 0.533737\, -0.501208 i \\
  & \text{$\lambda_{2} $=0.30} & 0.581318\, -0.0966934 i & 0.564485\, -0.293817 i & 0.534441\, -0.501349 i \\ \hline
 \text{$\lambda_{1} $=0.10} & \text{$\lambda_{2} $=0.05} & 0.578223\, -0.0965306 i & 0.561285\, -0.293355 i & 0.531053\, -0.500663 i \\
  & \text{$\lambda_{2} $=0.10} & 0.579966\, -0.0966228 i & 0.563088\, -0.293617 i & 0.532961\, -0.501052 i \\
  & \text{$\lambda_{2} $=0.15} & 0.581318\, -0.0966934 i & 0.564485\, -0.293817 i & 0.534441\, -0.501349 i \\
  & \text{$\lambda_{2} $=0.20} & 0.582466\, -0.0967527 i & 0.565673\, -0.293985 i & 0.535699\, -0.501597 i \\
  & \text{$\lambda_{2} $=0.25} & 0.583486\, -0.0968048 i & 0.566727\, -0.294133 i & 0.536815\, -0.501816 i \\
  & \text{$\lambda_{2} $=0.30} & 0.584413\, -0.0968519 i & 0.567686\, -0.294267 i & 0.53783\, -0.502012 i \\ \hline
 \text{$\lambda_{1} $=0.15} & \text{$\lambda_{2} $=0.05} & 0.579166\, -0.0965807 i & 0.56226\, -0.293497 i & 0.532086\, -0.500874 i \\
  & \text{$\lambda_{2} $=0.10} & 0.581318\, -0.0966934 i & 0.564485\, -0.293817 i & 0.534441\, -0.501349 i \\
  & \text{$\lambda_{2} $=0.15} & 0.582989\, -0.0967795 i & 0.566214\, -0.294061 i & 0.536271\, -0.50171 i \\
  & \text{$\lambda_{2} $=0.20} & 0.584413\, -0.0968519 i & 0.567686\, -0.294267 i & 0.53783\, -0.502012 i \\
  & \text{$\lambda_{2} $=0.25} & 0.585678\, -0.0969155 i & 0.568994\, -0.294447 i & 0.539215\, -0.502276 i \\
  & \text{$\lambda_{2} $=0.30} & 0.58683\, -0.0969729 i & 0.570185\, -0.294609 i & 0.540476\, -0.502514 i \\ \hline
 \text{$\lambda_{1} $=0.20} & \text{$\lambda_{2} $=0.05} & 0.579966\, -0.0966228 i & 0.563088\, -0.293617 i & 0.532961\, -0.501052 i \\
  & \text{$\lambda_{2} $=0.10} & 0.582466\, -0.0967527 i & 0.565673\, -0.293985 i & 0.535699\, -0.501597 i \\
  & \text{$\lambda_{2} $=0.15} & 0.584413\, -0.0968519 i & 0.567686\, -0.294267 i & 0.53783\, -0.502012 i \\
  & \text{$\lambda_{2} $=0.20} & 0.586073\, -0.0969353 i & 0.569402\, -0.294502 i & 0.539647\, -0.502358 i \\
  & \text{$\lambda_{2} $=0.25} & 0.58755\, -0.0970085 i & 0.57093\, -0.294709 i & 0.541264\, -0.502661 i \\
  & \text{$\lambda_{2} $=0.30} & 0.588898\, -0.0970745 i & 0.572323\, -0.294895 i & 0.542738\, -0.502933 i \\ \hline
 \text{$\lambda_{1} $=0.25} & \text{$\lambda_{2} $=0.05} & 0.580675\, -0.0966599 i & 0.56382\, -0.293722 i & 0.533737\, -0.501208 i \\
  & \text{$\lambda_{2} $=0.10} & 0.583486\, -0.0968048 i & 0.566727\, -0.294133 i & 0.536815\, -0.501816 i \\
  & \text{$\lambda_{2} $=0.15} & 0.585678\, -0.0969155 i & 0.568994\, -0.294447 i & 0.539215\, -0.502276 i \\
  & \text{$\lambda_{2} $=0.20} & 0.58755\, -0.0970085 i & 0.57093\, -0.294709 i & 0.541264\, -0.502661 i \\
  & \text{$\lambda_{2} $=0.25} & 0.589218\, -0.09709 i & 0.572655\, -0.294939 i & 0.543089\, -0.502997 i \\
  & \text{$\lambda_{2} $=0.30} & 0.590742\, -0.0971635 i & 0.57423\, -0.295146 i & 0.544756\, -0.503298 i \\ \hline
 \text{$\lambda_{1} $=0.30} & \text{$\lambda_{2} $=0.05} & 0.581318\, -0.0966934 i & 0.564485\, -0.293817 i & 0.534441\, -0.501349 i \\
  & \text{$\lambda_{2} $=0.10} & 0.584413\, -0.0968519 i & 0.567686\, -0.294267 i & 0.53783\, -0.502012 i \\
  & \text{$\lambda_{2} $=0.15} & 0.58683\, -0.0969729 i & 0.570185\, -0.294609 i & 0.540476\, -0.502514 i \\
  & \text{$\lambda_{2} $=0.20} & 0.588898\, -0.0970745 i & 0.572323\, -0.294895 i & 0.542738\, -0.502933 i \\
  & \text{$\lambda_{2} $=0.25} & 0.590742\, -0.0971635 i & 0.57423\, -0.295146 i & 0.544756\, -0.503298 i \\
  & \text{$\lambda_{2} $=0.30} & 0.592428\, -0.0972438 i & 0.575973\, -0.295372 i & 0.546599\, -0.503625 i \\
  \hline

\end{tabular}
\end{equation}

\subsection{Test of Stability for $\beta=\frac{3}{5}$ and $l=\frac{1}{2}$}

\begin{equation}    
    \begin{tabular}{|c|c|c|c|c|}
 \hline
 \text{$\lambda_{1}=\lambda_{2}$} & \text{WKB Order} & n=0 & n=1 & n=2 \\ \hline
 \text{$\lambda_{1}=\lambda_{2}=0.05$} & \text{Order $2$} & 0.190536\, - 0.120065 i & 0.197864 - 0.346854 i & 0.200694 - 0.797914 i \\
  & \text{Order $3$} & 0.178783 - 0.100374 i  & 0.14876 - 0.321383 i & 0.108962 - 0.546921 i \\
  & \text{Order $4$} & 0.18328 - 0.0979112 i & 0.149872 - 0.318999 i & 0.103055 - 0.578272 i \\
  & \text{Order $5$} & 0.182519 - 0.0964803 i & 0.146237 - 0.317308 i & 0.124872 - 0.582556 i \\
  & \text{Order $6$} & 0.184713 - 0.0953346 i & 0.147266 - 0.315089 i & 0.127799 - 0.569211 i \\
 \hline
  \text{$\lambda_{1}=\lambda_{2}=0.10$} & \text{Order $2$} & 0.192654 - 0.12002 i & 0.199827 - 0.347136 i & 0.20263 - 0.79878 i \\
  & \text{Order $3$} & 0.18121 - 0.10063 i & 0.151536 - 0.32177 i & 0.112237 - 0.547482 i \\
  & \text{Order $4$} & 0.185514 - 0.0982955 i & 0.152526 - 0.319683 i & 0.106188 - 0.578668 i \\
  & \text{Order $5$} & 0.184744 - 0.096835 i & 0.148895 - 0.317967 i & 0.127612 - 0.58298 i \\
  & \text{Order $6$} & 0.186869 - 0.0957341 i & 0.149904 - 0.315827 i & 0.130596 - 0.569662 i \\
  \hline
  \text{$\lambda_{1}=\lambda_{2}=0.15$} & \text{Order $2$} & 0.194864 - 0.119948 i & 0.201871 - 0.347354 i & 0.204643 - 0.799514 i \\
  & \text{Order $3$} & 0.183741 - 0.100876 i & 0.154435 - 0.322108 i & 0.115655 - 0.547956 i \\
  & \text{Order $4$} & 0.187848 - 0.0986704 i & 0.1553 - 0.320314 i & 0.109467 - 0.578929 i \\
  & \text{Order $5$} & 0.187072 - 0.0971849 i & 0.151692 - 0.31858 i & 0.130506 - 0.583274 i \\
  & \text{Order $6$} & 0.18912 - 0.096132 i & 0.152671 - 0.316537 i & 0.133538 - 0.570033 i \\
  \hline
   \text{$\lambda_{1}=\lambda_{2}=0.20$} & \text{Order $2$} & 0.197175 - 0.119844 i & 0.204002 - 0.3475 i & 0.20674 - 0.800096 i \\
  & \text{Order $3$} & 0.186382 - 0.101109 i & 0.157464 - 0.322388 i & 0.119224 - 0.548327 i \\
  & \text{Order $4$} & 0.190289 - 0.0990328 i & 0.158204 - 0.320879 i & 0.112902 - 0.579032 i \\
  & \text{Order $5$} & 0.18951 - 0.097528 i & 0.154639 - 0.319137 i & 0.133567 - 0.583414 i \\
  & \text{Order $6$} & 0.191476 - 0.0965267 i & 0.155577 - 0.317213 i & 0.136636 - 0.57031 i \\
  \hline
  \text{$\lambda_{1}=\lambda_{2}=0.25$} & \text{Order $2$} & 0.199594 - 0.119706 i & 0.206229 - 0.347562 i & 0.208927 - 0.800504 i \\
  & \text{Order $3$} & 0.189143 - 0.101326 i & 0.160631 - 0.322601 i & 0.122951 - 0.548581 i \\
  & \text{Order $4$} & 0.192848 - 0.0993793 i & 0.161248 - 0.321367 i & 0.116501 - 0.578949 i \\
  & \text{Order $5$} & 0.192071 - 0.0978621 i & 0.157752 - 0.319627 i & 0.13681 - 0.583375 i \\
  & \text{Order $6$} & 0.193945 - 0.0969164 i & 0.158635 - 0.317848 i & 0.139902 - 0.57048 i \\
  \hline
    \text{$\lambda_{1}=\lambda_{2}=0.30$} & \text{Order $2$} & 0.202131 - 0.119526 i & 0.208558 - 0.347529 i & 0.211212 - 0.800712 i \\
  & \text{Order $3$} & 0.192035 - 0.101523 i & 0.163947 - 0.322735 i & 0.12684 - 0.548696 i \\
  & \text{Order $4$} & 0.195536 - 0.0997055 i & 0.164443 - 0.32176 i & 0.120275 - 0.578646 i \\
  & \text{Order $5$} & 0.194764 - 0.0981844 i & 0.161047 - 0.320037 i & 0.140251 - 0.583126 i \\
  & \text{Order $6$} & 0.196537 - 0.097299 i & 0.161858 - 0.318434 i & 0.143348 - 0.570528 i \\
  \hline

\end{tabular}
\end{equation}

\end{document}